\documentclass[preprint,12pt]{elsarticle}

\usepackage{geometry}
 \usepackage[dvipsnames]{xcolor}
 
\usepackage{amssymb}
\usepackage{amsmath}
\usepackage{multirow}

\newcommand{\review}[1]{\color{black}#1}

\makeatletter
\def\ps@pprintTitle{
  \let\@oddhead\@empty
  \let\@evenhead\@empty
  \let\@oddfoot\@empty
  \let\@evenfoot\@oddfoot
}
\makeatother

\begin{document}

\begin{frontmatter}



\title{Adaptive interface-Mesh un-Refinement (AiMuR) based Sharp-Interface Level-Set-Method for
 Two-Phase Flow}



\author[1,2]{Kuntal Patel}
\author[2]{Javed Shaikh}
\author[1]{Absar Lakdawala}
\author[2]{Atul Sharma \corref{cor}}
\ead{atulsharma@iitb.ac.in}
\address[1]{Department of Mechanical Engineering, Nirma University, Ahmedabad, India.}
\address[2]{Department of Mechanical Engineering, Indian Institute of Technology Bombay, Mumbai, India.}
\cortext[cor]{Corresponding author}

\begin{abstract}
\noindent
Adaptive interface-Mesh un-Refinement (AiMuR) based Sharp-Interface Level-Set-Method (SI-LSM) is proposed for both uniform and non-uniform Cartesian-Grid. The AiMuR involves interface location based dynamic un-refinement (with merging of the four control volumes) of the Cartesian grid away from the interface. The un-refinement is proposed for the interface solver only. A detailed numerical methodology is presented for the AiMuR and ghost-fluid method based SI-LSM. Advantage of the novel as compared to the traditional SI-LSM is demonstrated with a detailed qualitative as well as quantitative performance study, involving the SI-LSMs on both coarse grid and fine grid, for three sufficiently different two-phase flow problems: dam break, breakup of a liquid jet and drop coalescence. A superior performance of AiMuR based SI-LSM is demonstrated - the AiMuR on a coarser non-uniform grid ($NU_{c}^{AiMuR}$) is almost as accurate as the traditional SI-LSM on a uniform fine grid ($U_{f}$) and takes a computational time almost same as that by the traditional SI-LSM on a uniform coarse grid ($U_{c}$). \review{The AMuR is different from the existing
Adaptive Mesh Refinement (AMR) as the former involves only mesh un-refinement while the later involves both refinement and un-refinement of the mesh.} Moreover, the proposed computational development is significant since the present adaptive un-refinement strategy is much simpler to implement as compared to that for the commonly used adaptive refinement strategies. The proposed numerical development can be extended to various other multi-physics, multi-disciplinary and multi-scale problems involving interfaces.   

\end{abstract}




\end{frontmatter}





\section{Introduction}\label{intro}

Computational Fluid Dynamics (CFD) is a theoretical-method of scientific
and engineering investigation, concerned with the development and
application of a video-camera like tool (a software) which is used
for a unified cause-and-effect based analysis of a fluid-dynamics
as well as heat and mass transfer problem; presented in a recently
published book on CFD by Sharma \cite{Sharma2017}.  He proposed a conservation
law based finite volume method for a discrete (independent of continuous)
mathematics based derivation of the system of algebraic equations that
are the governing equations in CFD.   The CFD for a multi-fluid flow
is commonly called as Computational Multi-Fluid Dynamics (CMFD) that
involves the application of the conservation laws to each  of the fluids
in the multi-fluid system. A key difference between  CFD and CMFD
 is the lower dimensional fluid-fluid interface that separates two fluids.
In order to track/capture the interface, various CMFD methodologies
are available in the literature. Reliability of any CMFD methodology
depends upon  its ability to handle (a) the jump in thermo-physical
properties across the interface and (b) the severe changes in the
topology of the interface. Thus, a CMFD methodology demands a high level
of grid resolution $-$ especially near the interface $-$ to achieve
desired numerical accuracy. However, there must be a trade-off between
the selection of a grid strategy (for better numerical accuracy) and
the associated computational cost/time.

Front Tracking Method (FTM) (Juric and Tryggvason \cite{Juric1996})
is a CMFD method that belongs to the class of Lagrangian framework,
wherein the interface is tracked explicitly by the means of markers. However,
it demands some additional modeling in order to simulate the merger/breakup
of interfaces. Other CMFD methodologies like Volume of Fluid (VOF)
and Level Set Method (LSM) follow  the Eulerian framework. In VOF
and LSM, an additional scalar field is defined which helps in capturing
interface implicitly. Volume of Fluid (Hirt and Nicholas \cite{Hirt1981})
is one of the widely adopted multi-fluid methodologies where an interface
is defined by a scalar field  based on a volume fraction. Level Set Method (Osher
and Sethian \cite{Osher1988}, Sussman et al. \cite{sussman1994}) is another
interface capturing technique wherein an interface is represented
by a level set function $\phi=0$, where $\phi$ is a scalar field
defined as a signed normal distance. Implementation of surface tension
is very straightforward in LSM as the geometrical parameters can be
obtained directly with the help of the normal distance function field
for $\phi$. Detailed literature survey on level set method based
developments and applications can be found in a recent review-papers
by Sharma \cite{shrma2015} and Gibou et al. \cite{gibou2018}. Broadly,
there are two types of LSMs: Diffuse Interface Level Set Method (DI-LSM)
\cite{sussman1994} and ghost fluid method based Sharp Interface Level
Set method (SI-LSM) \cite{fedkiw1999}. The interfacial force due to surface
tension is modeled as a body-force in the DI-LSM while the SI-LSM
considers the force as the more realistic surface-force acting at
the interface (implemented as an interfacial boundary condition for
pressure-jump across the interface) \cite{shaikh2018}. The SI-LSM as compared to the DI-LSM leads to a substantial reduction in mass error \cite{shaikh2018} $-$ the biggest disadvantage of a LSM. Detrixhe and Aslam \cite{detrixhe2015} introduced an algorithm to obtain a volume fraction field from a level set function, or vice versa, with the second-order accuracy for interface location and first-order accuracy for interface normal. This algorithm can be employed to combine the respective advantages of VOF and LSM.


For most of the multi-fluid flow problems, the fluid-dynamics actions
are concentrated in the vicinity of the lower dimensional fluid-fluid
interface. The interfacial physics demands sufficiently large grid
resolution near the interface for an accurate numerical solution.
An efficient grid strategy in CMFD should generate large grid resolution
near the interfacial region as compared to far-away  from the interface.
Based on this consideration, CMFD researchers have worked on the development
and implementation of computationally effective grid strategies such
as stretched/clustered non-uniform grid, nested block grid, and adaptive
mesh. Using the non-uniform grid for FTM, Thomas et al. \cite{thomas2010}
studied thin-film flows during the impact of droplets in an inclined
channel. Furthermore, using the non-uniform grid, a VOF based multi-fluid
computations was presented by several researchers: Richards et al.
\cite{richards1995} studied a jet-breakup problem, Kobayashi et al. \cite{kobayashi2004}
studied formation of emulsion droplet in micro-channels, Yanke et
al. \cite{Yanke2015} studied a electroslag remelting problem,  Koukouvinis
et al. \cite{koukouvinis2016} studied bubble collapse and expansion near
the free surface, Waters et al. \cite{waters2017} studied breakup of
turbulent sprays,  and Sultana et al. \cite{sultana2017} incorporated
phase change process to study  droplet dynamics.   Application of
the non-uniform grid for LSM was presented by a few researchers: Jarrahbashi and Sirignano
\cite{Jarrahbashi2014} for simulation of liquid-injection at high pressure,
Montazeri \& Ward \cite{Montezari2014} for proposing a numerical methodology
for generalized body force, and Vilegas et al. \cite{villegas2016,villegas2017}
for simulating leidenfrost effect. Finally, application of the non-uniform
grid for a Coupled Level Set and Volume of Fluid (CLSVOF) method was
presented by Ferrari et al. \cite{ferrari2017} for simulation of micro-scale
multi-phase flows.

\review{Another class of efficient grid strategy based numerical method is Adaptive Mesh Refinement (AMR) \cite{berger1984}  which involves dynamic refinement as well as un-refinement of the grid that is based on certain predefined criteria.} The AMR based
VOF method was presented by Popinet \cite{popinet2009} and AMR based LSM
was presented by Sussman et al. \cite{sussman1999} for incompressible
multi-phase flows.  Whereas, for compressible multi-phase flow, AMR
based LSM was presented by Nourgaliev and Theofanous \cite{nourgaliev2007}.
Implementation of AMR can be done using Quadtree and Octree data structures (Samet
\cite{samet1989,samet1990}) for 2D and 3D problems, respectively. Brun
et al. \cite{brun2012} used Hash Table  instead of Quadtree data
structure with a local LSM  \cite{adalsteinsson1995,peng1999}. In VOF framework, Theodorakakos and Bergeles \cite{theodorakakos2004} proposed adaptive mesh refinement of the interfacial cartesian grid by treating it as an unstructured mesh (\emph{i.e.} a computational cell can possess an arbitrary number of faces and neighboring cells). Recently, Antepara et al. \cite{antepara2019} studied the path instability of rising bubbles at high Reynolds number by integrating the conservative level-set method with their earlier AMR framework \cite{antepara2015} for single-phase turbulent flows. \review{Using separate grids for flow solver and interface solver, Herrmann \cite{hermann2008} proposed a Refined Level Set Grid (RLSG) Method with the purpose of simulating two-phase flows with a high-density ratio.} Using twice the number
of grid for the interface as compared to the grid for flow solver, a dual-grid
approach was proposed in our research group for DI-LSM \cite{gada2011}
that was recently extended for the SI-LSM by Shaikh et al. \cite{shaikh2019}.

Summary of the literature review, based on the various types of grid
structure and CMFD methodology, are presented in  Table \ref{tab:Summary-of-literature}.
The table shows that although there is some work on adaptive mesh
refinement (AMR) for VOF method and LSM, no such work is found for
Adaptive Mesh \emph{un-Refinement} (AMuR) which is proposed in the
present work. \review{Our AMuR can be considered as a variant of the AMR, with only mesh un-refinement in the AMuR and both refinement and un-refinement of the mesh in the AMR.} The motivation for the proposition of the AMuR as compared
to the already existing AMR is the ease in the implementation of the
AMuR since it allows the usage of commonly used matrix as the \emph{structured}
data structure (with in-built neighboring information) as compared
to the AMR that requires tree-based \emph{unstructured} data structure.
Moreover, as compared to the AMR presented in the Table \ref{tab:Summary-of-literature}
which uses adaptive refinement for both flow solver and interface
solver, the AMuR proposed here considers adaptive unrefinement (of the first level) only
for the interface solver $-$ called as Adaptive interface Mesh un-Refinement
(AiMuR). Since the value of the level set function (representing the
interface) is numerically relevant only up to certain distance away
from the interface, the unrefinement is proposed away from the interface
as merging of the four Cartesian control volumes for level set function.
The motivation is to obtain almost the same accuracy, with a substantial
reduction in the computational time for the solution of level set
equations based interface solver by the AiMuR as compared to uniform/non-uniform Cartesian grid. The objective of this work is to present a detailed \textcolor{black}{CMFD methodology for AiMuR based SI-LSM (section \ref{sec:Concept-of-Semi-Adaptive} and \ref{sec:method}) (on both uniform as well as non-uniform Cartesian grid) and performance study for the AiMuR (section \ref{sec:validation}) (as compared to the results obtained without the unrefinement) on three different two-fluid flow problems: dam break simulation, }breakup of a liquid jet and drop coalescence\textcolor{black}{.}

\begin{table}
\protect\caption{\label{tab:Summary-of-literature}Summary of literature review and the present work based
on the \review{mesh type/algorithm} and computational multi-fluid dynamics (CMFD) methodology. }

\centering{}%
\begin{tabular}{|l|c|c|}
\hline
\multirow{2}{*}{Published Work} & \review{Mesh} & CMFD \tabularnewline
 & \review{Type/Algorithm} & Methodology\tabularnewline
\hline
Thomas et al. \cite{thomas2010} & \multirow{2}{*}{Non-uniform grid} & \multirow{2}{*}{VOF}\tabularnewline
\cline{1-1}
Richards et al. \cite{richards1995} &  & \tabularnewline
\cline{1-1}
Kobayashi et al. \cite{kobayashi2004} &  & \tabularnewline
\cline{1-1}
Yanke et al. \cite{Yanke2015} &  & \tabularnewline
\cline{1-1}
Koukouvinis et al. \cite{koukouvinis2016} &  & \tabularnewline
\cline{1-1}
Waters et al. \cite{waters2017} &  & \tabularnewline
\cline{1-1}
Sultana et al. \cite{sultana2017} &  & \tabularnewline
\cline{1-1} \cline{3-3}
Jarrahbashi and Sirignano \cite{Jarrahbashi2014} &  & \multirow{2}{*}{LSM}\tabularnewline
\cline{1-1}
Montazeri and Ward \cite{Montezari2014} &  & \tabularnewline
\cline{1-1}
Vilegas et al. \cite{villegas2016,villegas2017} &  & \tabularnewline
\cline{1-1} \cline{3-3}
Ferrai et al. \cite{ferrari2017} &  & CLSVOF\tabularnewline
\hline
Popinet \cite{popinet2009}  & \multirow{2}{*}{AMR} & \multirow{2}{*}{VOF}\tabularnewline
\cline{1-1}
Theodorakakos and Bergeles \cite{theodorakakos2004} &  & \tabularnewline
\cline{1-1}
\cline{3-3}
Sussman et al. \cite{sussman1999} &  & \multirow{2}{*}{LSM}\tabularnewline
\cline{1-1}
Nourgaliev and Theofanous \cite{nourgaliev2007} &  & \tabularnewline
\cline{1-1}
Antepara et al. \cite{antepara2019} &  & \tabularnewline
\cline{1-2}
Hermann \cite{hermann2008} & RLSG & \tabularnewline
\cline{1-2}
Gada and Sharma \cite{gada2011} and Shaikh et al. \cite{shaikh2019} & DGLSM & \tabularnewline
\cline{1-2}
Present work & AiMuR & \tabularnewline
\hline
\end{tabular}
\end{table}

\section{Ghost Fluid Method based Sharp-Interface Level Set Method}\label{sec3}
\begin{figure}
\begin{centering}
\includegraphics[width=14cm]{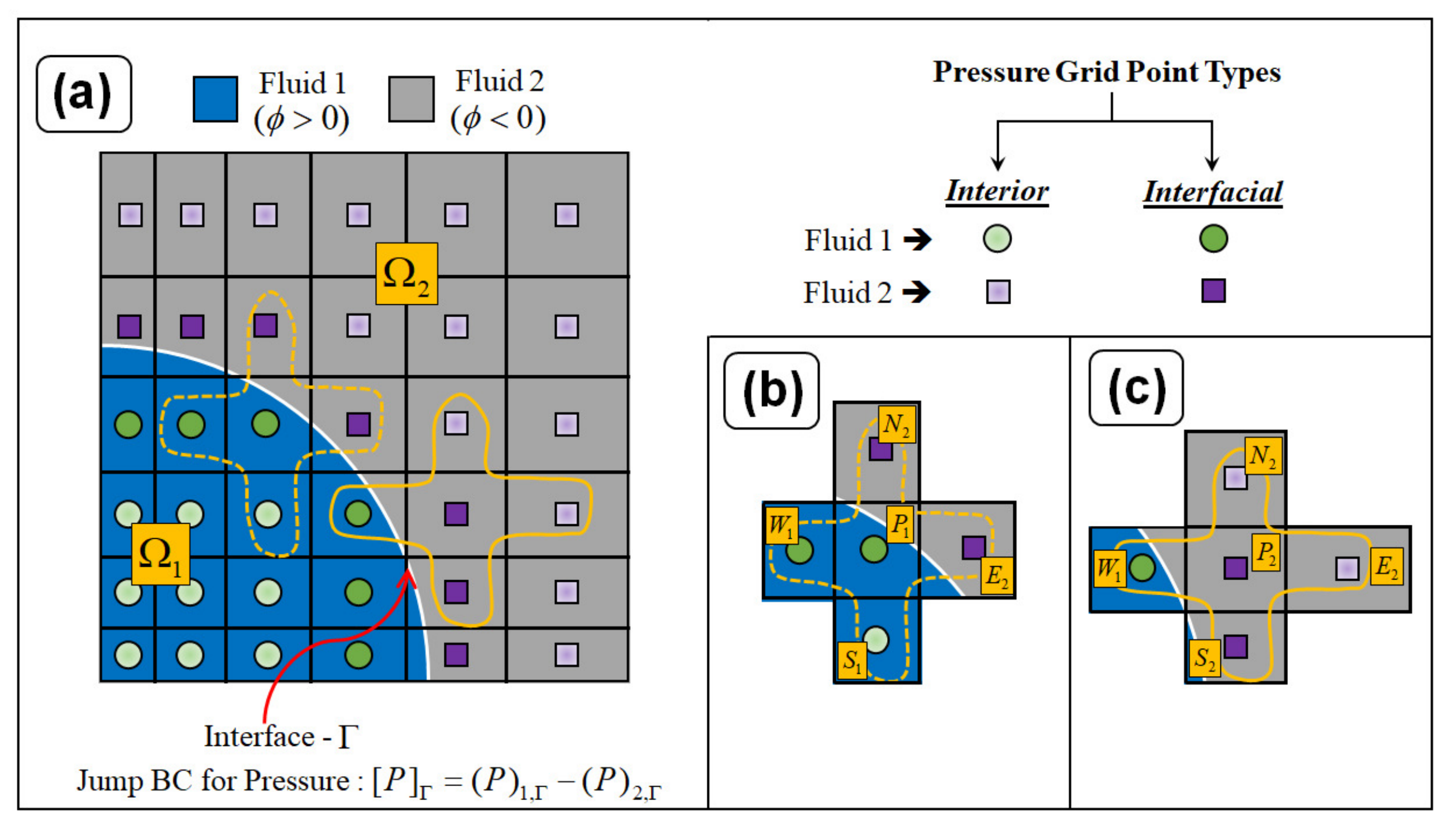}
\par\end{centering}
\protect\caption{\label{fig:Computational-domain-with}($a$) A representative computational
domain along with two-fluid interface and non-uniform grid distribution
and  of different types of pressure grid points for the two-fluid
sub-domains ($\Omega_{1}$ and $\Omega_{2}$). Mixing of pressure
grid point (from another fluid sub-domain) during the solution of
pressure Poisson equation for interfacial pressure grid points is
shown  for $\Omega_{1}$ in ($b$) and for $\Omega_{2}$ in ($c$). }
\end{figure}

In two-phase flows, the interface $\Gamma$ is considered as sharp
curve separating the two disjoints $\left(\Omega\equiv\Omega_{1}\cup\Omega_{2}\right)$
as shown in Fig. \textcolor{black}{\ref{fig:Computational-domain-with}}, with $\phi>0$ in fluid-1 and $\phi<0$ in fluid-2.
For the SI-LSM based simulation of two-phase flow, the incompressible
Navier-Stokes (continuity and momentum) equations (for both the fluids)
are solved for the spatial and temporal variation of the flow field
$-$ with an interfacial boundary condition, implemented using Ghost
Fluid method \cite{fedkiw1999}. The unsteady flow field is used to obtain
the temporal variation of the interface.

\subsection{\review{Single Fluid Formulation}}

For obtaining a single velocity and pressure field for both the fluids
in a two-fluid flow, a single field formulation based governing equations
and interfacial boundary conditions for SI-LSM are presented in a
recent work from our research group; separately for two-phase flow
without \cite{shaikh2018} and with \cite{shaikh2019} phase change. For
two-phase flow without phase change considered here, the various functions
used in a LSM and the formulation for sharp as well as diffuse interface
LSM are presented in-detail by Shaikh et al. \cite{shaikh2018}; thus,
the formulation is presented concisely in separate subsections below.

For the mathematical formulation, it is assumed that the interface
is massless with zero-thickness, and no-slip in tangential velocity.
Constant material properties are considered, but not equal for each
phase, \emph{i.e.}, the bulk fluids are incompressible. The surface
tension coefficient is assumed to be constant, and its tangential
variation along the interface is neglected. The effects of radiation,
viscous dissipation, and energy contribution due to interface stretching
are neglected.

\subsubsection{Governing Equations for Two-Fluid Flow Properties}\label{sec2.1.1}

Non-dimensional form of the conservation equations for the two-fluid
flow (Navier-Stokes equations) are given as

\noindent \textbf{Volume-Conservation (Continuity) Equation:}
\begin{equation}
\nabla\cdot\mathbf{U}=0,\label{eq:conti}
\end{equation}

\noindent \textbf{Momentum-Conservation Equation:}
\begin{equation}
\frac{\partial\mathbf{U}}{\partial\tau}+\nabla\cdot\left(\mathbf{U}\mathbf{U}\right)=-\frac{\nabla P}{\chi_{i}}+\frac{1}{\chi_{i}Re}\nabla\cdot\left(2\eta_{i}\mathbf{D}\right)-\frac{1}{Fr^{2}}\hat{j},\label{eq:Mom}
\end{equation}

\noindent \textcolor{black}{where rate of deformation tensor, $\mathbf{D}=0.5\left[\nabla\mathbf{U}+\left(\nabla\mathbf{U}\right)^{T}\right]$.
Furthermore, $\chi_{i}$ and $\eta_{i}$ are the non-dimensional density
and viscosity; calculated using a sharp Heaviside function }\cite{shaikh2018}\textcolor{black}{.
Also $\hat{j}$ is the unit vector for gravity ($\hat{j}=<0,-1>$).
}Using characteristic scales as $l_{c}$ for length and $u_{c}$ for
velocity, the non-dimensional spatial as well as temporal coordinates,
non-dimensional flow properties, and non-dimensional governing parameters
(Reynolds number $Re$ and Froude number $Fr$), the non-dimensional variables in the above equations
are defined as

\[
\mathbf{X}=\frac{x}{l_{c}},\,\mathbf{U}=\frac{u}{u_{c}},\,\tau=\frac{tu_{c}}{l_{c}},\: P=\frac{p}{\rho_{1}u_{c}^{2}},\: Re=\frac{\rho_{1}u_{c}l_{c}}{\mu_{1}}\mbox{ and }Fr=\frac{u_{c}}{\sqrt{gl_{c}}}.
\]

For the two-phase as compared to most of the single-phase flow, note
that the above momentum equations consist of gravity force as the
additional force while the force due to surface tension which also
appears for a two-phase flow is incorporated in the SI-LSM as an interfacial
boundary condition during the solution of pressure Poisson equation
(obtained from the above volume conservation equation); presented
below. Also note that the force due to surface tension is considered as
a sharp surface force in the SI-LSM \cite{shaikh2018} while it is modeled
as a volumetric force term (within the thickness of the diffused interface)
in the above momentum equation for the DI-LSM \cite{gada2011}.

\subsubsection{Governing Equations for Two-Fluid Interface}

Physically, the two-fluid interface is advected by the fluid-flow;
obtained by solving the governing equations in the previous subsection.
Mathematically, in a LSM, the unsteady interface advection is represented
by an advection equation for a signed normal distance function representing
the interface, \emph{i.e.}, level set function $\phi$. However, after
the advection, the interface representing $\phi$ no more remains
as a normal distance function and another equation called as reinitialization
equations is solved using the pseudo transient approach. The reinitialization
is essential for an accurate calculation of surface tension, jump
terms, and thermophysical properties. Thus, the governing equations
for the two-fluid interface are given as

\noindent \textbf{Level-Set Advection (Mass-Conservation) Equation}:
\begin{equation}
\frac{\partial\phi}{\partial\tau}+\mathbf{U}_{a}\cdot\nabla\phi=0,\thinspace\thinspace\thinspace\thinspace\thinspace\thinspace\thinspace\label{eq:lsadv}
\end{equation}
where $\mathbf{U}_{a}$ is the advection velocity which is equal to
the bulk-velocity $\mathbf{U}$ (obtained from the solution of the
above volume and momentum conservation equations). The above equation
was derived from a mass conservation equation by Gada and Sharma \cite{gada2009}.

\noindent \textbf{Reinitialization Equation}:

\begin{equation}
\frac{\partial\phi}{\partial\tau_{s}}+S\left(\phi_{o}\right)\widehat{n}\cdot\nabla\phi-S\left(\phi_{o}\right)=0,\label{eq:2.21}
\end{equation}

\noindent where $S\left(\phi_{o}\right)$ is a sign function and $\tau_{s}$
is a pseudo time step. After getting the converged solution of Eq.~(\ref{eq:2.21}),
level set field will again become signed normal distance with respect
to zero level set value. Here, partial differential equation based
reinitialization (Sussman et al. \cite{sussman1994}) is used.

\subsubsection{Interfacial Boundary Conditions}

In a computational fluid dynamics (CFD), the unsteady velocity field
is obtained from the momentum conservation equation (Eq.~(\ref{eq:Mom}))
and the pressure field is left with the volume conservation equation
(Eq.~(\ref{eq:conti})) which does not consist of any pressure term
\cite{Sharma2017}; thus, a predictor-corrector method is used to convert
the volume conservation equation into a pressure Poisson equation
in a pressure projection method. While solving the pressure Poisson
equation and the momentum equation for a two-phase flow in a single
field formulation based SI-LSM, interfacial boundary conditions along
with the boundary conditions for the flow properties at the boundary
of the domain are required. The interfacial BCs involve jump boundary
conditions for pressure and velocity at the interface $\Gamma$; given
as

\begin{eqnarray}
\left[P\right]_{\Gamma} & = & \frac{2}{Re}\left[\eta\right]\hat{N\centerdot}\left(\nabla U\centerdot\hat{N},\nabla V\centerdot\hat{N}\right)+\frac{\kappa}{We},\label{eq:Jump_P_Phase-1}\\
{}[\mathbf{U}]_{\Gamma} & = & 0,\nonumber
\end{eqnarray}

\noindent where the above equation for pressure is obtained by applying
Newton's II law of motion at the interface $\Gamma$ and a detailed
derivation of the pressure-jump BC was presented by Shaikh et al.
\cite{shaikh2018}. The interfacial force balance considers the force
due to surface tension as the \emph{sharp surface-force} which is
balanced with both pressure force and normal viscous force in viscous
flow across the interface\textcolor{blue}{{} }\textcolor{black}{\cite{francois2006}}.

The interfacial boundary condition across the interface
is shown in Fig. \ref{fig:Computational-domain-with}. As shown in
the figure, for pressure, the jump condition notation across the interface
is expressed as\textcolor{black}{{} ${\color{black}{\color{red}{\color{black}[\cdot]_{\Gamma}=\left(\cdot\right)_{1,\Gamma}-\left(\cdot\right)_{2,\Gamma}}}}$};
$\left(\cdot\right)_{1,\Gamma}$ and $\left(\cdot\right)_{2,\Gamma}$
are the value of the flow properties at the interface $\Gamma$ from
the heavier and lighter phase in the $\Omega_{1}$ and $\Omega_{2}$
region, respectively.

\section{Adaptive interface-Mesh un-Refinement}\label{sec:Concept-of-Semi-Adaptive}

\begin{figure}
\begin{centering}
\includegraphics[scale=0.5]{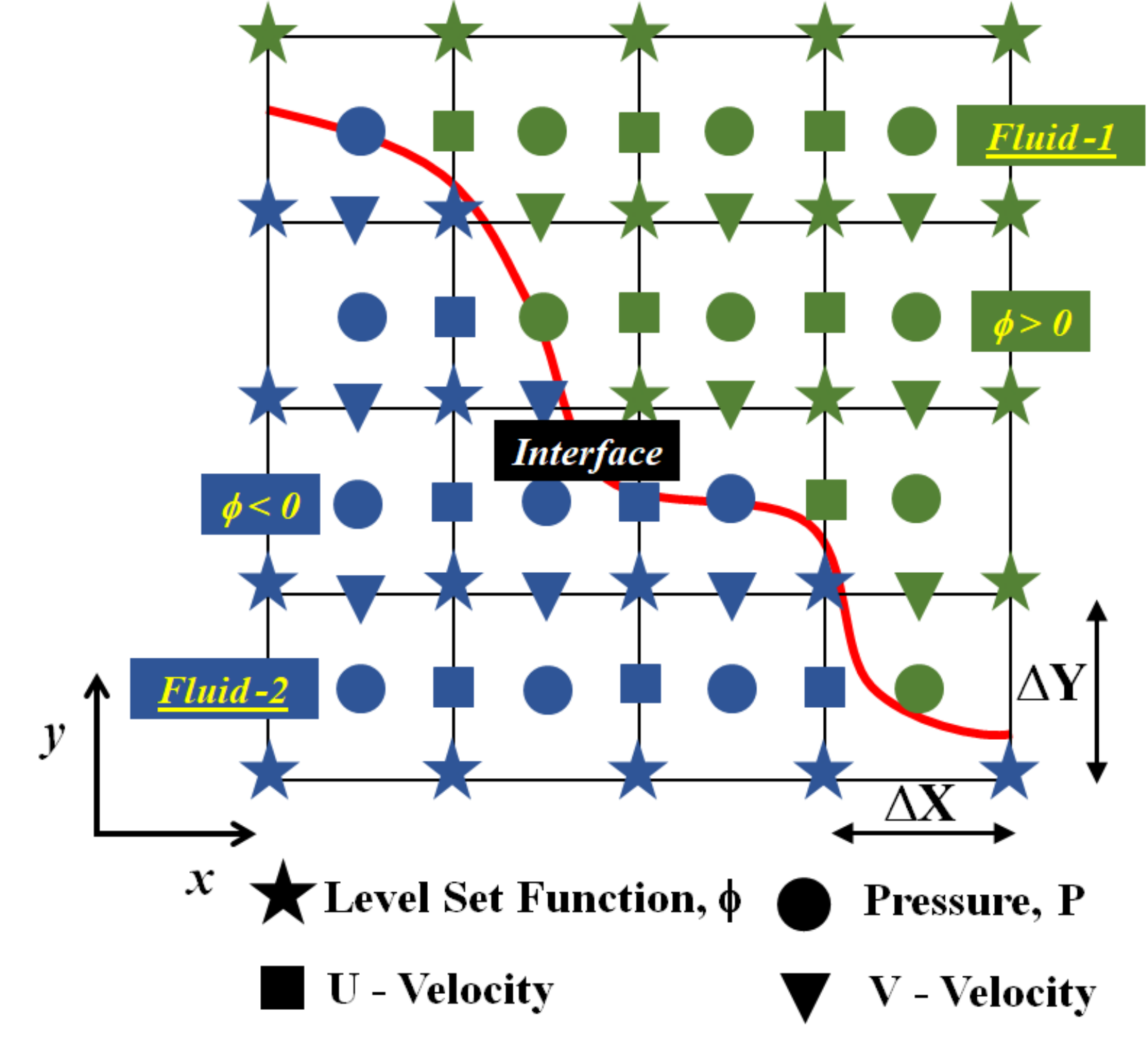}
\par\end{centering}

\protect\caption{\label{fig:Computational-stencil-with} A representative 2D Cartesian
grid along with the \textit{staggered} grid points for flow-properties
and level set function $\phi$ in a two-fluid problem.}
\end{figure}

\begin{figure}
\begin{centering}

\par\end{centering}

\noindent \begin{centering}
\includegraphics[width=14cm]{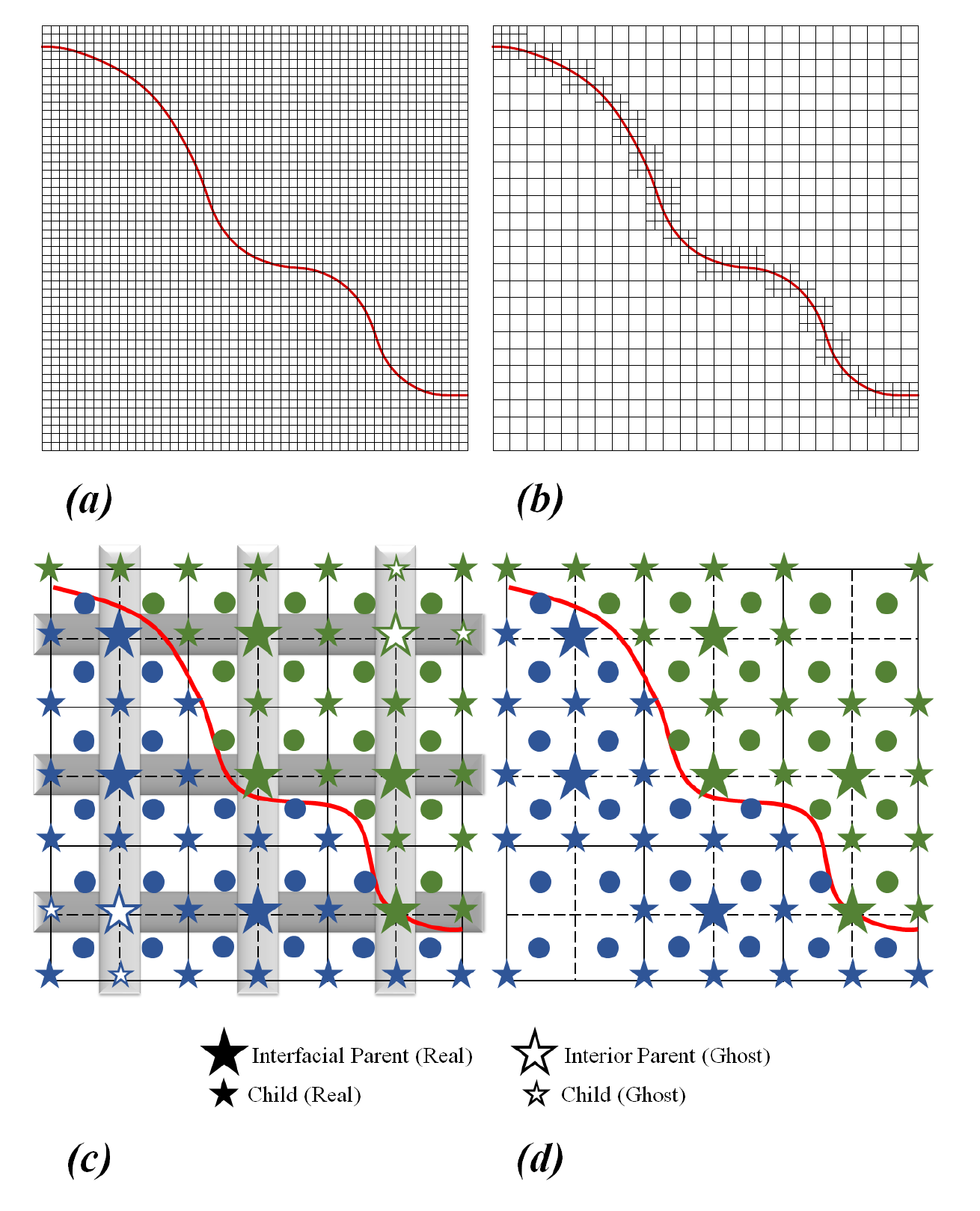}
\par\end{centering}

\protect\caption{\label{fig:(a)-Distribution-of}\review{A representative Cartesian grid for a two-fluid system with an interface: ($a,c$) initial uniform grid at $\tau=0$  for the interface and at all time instants for the flow and ($b,d$) dynamically unrefined grid. Here, ($c$) and ($d$) show the various types of grid points considered to implement the instantaneous interface based dynamic mesh un-refinement. The unrefined interface grid in ($d$) is obtained from the uniform grid for the flow-properties in ($c$) after eliminating ghost-parent and ghost-child level set grid points. Note that a narrow band of the fine grid (at the interface) is shown in ($b$) for representative purpose only and a much wider band is used in the present method for accurate computations.}}
\end{figure}

For the staggered grid used here, the grid points  for pressure, velocity
and level set function are shown in Fig.~\ref{fig:Computational-stencil-with}.
 For the implementation of the interface-mesh unrefinement,  all level
set grid points are tagged as a parent or a child grid point. All parent level set grid points are further tagged as interfacial or interior grid points. Finally, each parent and child grid point is tagged as real or ghost grid point. The various types of level set grid points are shown in
Fig.~\ref{fig:(a)-Distribution-of}. The figure also shows an interface
representing the two-fluid. A representative 2D Cartesian grid is shown in Fig.~\ref{fig:(a)-Distribution-of}($a$)
as a uniform mesh for solving the Navier-Stokes equations (Eq.~(\ref{eq:conti})
and (\ref{eq:Mom})) along with the interfacial boundary conditions
(Eq.~(\ref{eq:Jump_P_Phase-1})); and Fig. \ref{fig:(a)-Distribution-of}($b$)
as the adaptive unrefined interface mesh for solving the level set equations
(Eq.~(\ref{eq:lsadv}) and (\ref{eq:2.21})). The solution of the
respective set of equations results in unsteady flow properties (U,V,
and P) and level set function $\phi$. Fig.~\ref{fig:(a)-Distribution-of}($b$)
shows a merging of the four finer control volumes (CVs) in Fig. \ref{fig:(a)-Distribution-of}($a$) to a coarser control volume, for those CVs which are slightly away
from the interface. This results in the unrefinement of the interface
mesh which is away from the interface and the unrefinement is time-wise adaptive to the position of the interface which evolves with
time $-$ called here as Adaptive interface-Mesh un-Refinement (AiMuR).

The implementation and algorithmic details for the adaptive unrefinement
of the interface mesh (Fig.~\ref{fig:(a)-Distribution-of}($b$)) $-$
from the fixed flow-properties mesh (Fig.~\ref{fig:(a)-Distribution-of}($a$))
$-$ are presented with the help of Fig.~\ref{fig:(a)-Distribution-of}($c$)
and \ref{fig:(a)-Distribution-of}($d$). Fig.~\ref{fig:(a)-Distribution-of}($c$)
shows a tagging of each level set grid point as parent or child, interfacial
or interior, and real or ghost grid points; and Fig.~\ref{fig:(a)-Distribution-of}($d$)
shows only real (not the ghost) grid points that lead to the AiMuR.
The three types of tagging for each level set function grid point
are as follows:

\noindent 1. \textit{Tag as parent or child grid point}: parent if
both the running indices \textit{i} and \textit{j} are even; otherwise,
child.

\noindent 2. \textit{Tag parent grid points as interfacial or interior}:
all parent level set grid points with at least one adjoining neighbor
(east/west/north/south) parent level set grid point in another fluid
are tagged as an interfacial parent grid point. Identification of
the interfacial and non-interfacial parent grid points are done by
considering change in the sign of level set function $-$  interfacial
parent grid point if the product of level set function at parent grid
point $\phi_{P}$ and at any of the adjoining parent neighbor $\phi_{NB}$
is negative ($\phi_{P}\phi_{NB=W,E,S,N}<0$); otherwise, consider
the grid point as interior parent grid point.

\noindent 3. \textit{Tag as ghost or real}: interior as ghost and
interfacial as real, for the parent grid points; whereas, for the
child grid points, the adjoining neighbor (north, south, east, and
west) child grid points of a ghost interior parent grid point are
considered as ghost and all the other child nodes in the domain are
considered as real. Note that the common adjoining child neighbors
of ghost interior parent grid points and real interfacial parent grid
points are considered as real child grid points.

Based on the proposed definition of parent and child level set grid
points and interface configuration shown in Fig.~\ref{fig:(a)-Distribution-of}($c$),
level set function grid points at the intersection of the  horizontal strips
and vertical strips (marked in the figure)  are the parent level set grid points. Here, words
real and ghost are used in the sense that level set equations are
solved only for real grid points and not for ghost grid points. This
classification of parent and child level set grid points into real
and ghost grid points  generates  level set grid point distribution
as finer in the interfacial region and coarser  in the non-interfacial
region. Implementing the  tagging procedure (discussed above) for
level set grid point distribution shown in Fig.~\ref{fig:(a)-Distribution-of}($c$)
and then eliminating the ghost parent and child nodes results in real
grid points distribution  as shown in Fig.~\ref{fig:(a)-Distribution-of}($d$). 

The above-discussed implementation  results in a \textit{single level}
 adaptive interface mesh unrefinement. However, the present unstructured
adaptive grid-like distribution is implicitly achieved by the unrefinement
using tagging and without involving any tree data structure. Once
the unrefinement is done, the level set grid will have the same resolution
as that of flow grid in the interfacial region while the level set
grid away from interface will be coarser by the single-level.  Limiting
the refined grid close to the interface is justified since the value
of the level set function $\phi$ close to the physically relevant
interface $\phi=0$ is only numerically relevant $-$ $\phi$ values
close to the interface are only involved in the calculation of interfacial
parameters such as fluid properties, curvature, and jump in the flow
properties.

Although the above implementation details for AiMuR are presented in Fig. \ref{fig:(a)-Distribution-of}
for one interfacial cell on each side of the interface, note that
every interfacial parent level set grid point and its three neighbor
parent level set grid points in all the four directions (east/west/north/south)
are considered in the present work for a more accurate CMFD computations.
Thus, the wider interfacial band is considered in the proposed AiMuR since
the one interfacial cell-based AiMuR shown in Fig.~\ref{fig:(a)-Distribution-of}($b$)
 leads to an inaccurate solution. Moreover, as commonly used in AMR
\cite{popinet2009,theodorakakos2004} and used here for a more efficient AiMuR, the adaption of the grid is
done after certain number of time steps (instead of after every $\Delta\tau$); $50\Delta\tau$, $150\Delta\tau$, and $15\Delta\tau$ are considered for the dam break simulation, liquid jet breakup, and droplet coalescence problems (presented below), respectively. \review{These time-periods of the unrefinment are obtained after an unrefinment time-period independence study, presented below in subsection~\ref{sub:Unref_freq}.} The periodic grid unrefinement  justifies the need for the wider
interfacial band of the finer grid. Further, it is ensured that the interface
does not enter into unrefined region during the above mentioned time
interval of the periodic unrefinement for the three problems considered in the present work. 

\section{Numerical Methodology}\label{sec:method}

\subsection{Solution of Volume and Momentum Conservation Equations}

Numerical solution of volume and momentum conservation equations is
obtained by the pressure projection method  in the present study. Here,
semi-implicit formulation is adopted wherein the volume conservation
equation is treated implicitly and all the terms of the momentum conservation
equation  except advection term are treated implicitly.
Temporal discretization of the corresponding equations (Eq.~(\ref{eq:conti})
and (\ref{eq:Mom})) are given as

\begin{equation}
\nabla\cdot\mathbf{U}_{P}^{n+1}=0,\label{eq:volume conservation}
\end{equation}

\begin{equation}
\frac{\mathbf{U}_{P}^{n+1}-\mathbf{U}_{P}^{n}}{\Delta\tau}+\nabla\cdot(\mathbf{U}_{P}^{n}\mathbf{U}_{P}^{n})=-\frac{\nabla P^{n+1}}{\chi_{i}^{n}}+\frac{1}{\chi_{i}^{n}Re}\nabla\cdot(2\eta_{i}^{n}\mathbf{D}^{n+1})-\frac{1}{Fr^{2}}\hat{j}.\label{eq:6}
\end{equation}

\subsubsection{Semi-Implicit Pressure Projection Method}

In the pressure projection method,  velocity field  is predicted by
neglecting the pressure term in Eq.~(\ref{eq:6}); given as

\begin{equation}
\frac{\mathbf{U}_{P}^{*}-\mathbf{U}_{P}^{n}}{\Delta\tau}+\nabla\cdot(\mathbf{U}_{P}^{n}\mathbf{U}_{P}^{n})=\frac{1}{\chi_{i}^{n}Re}\nabla\cdot(2\eta_{i}^{n}\mathbf{D}^{*})-\frac{1}{Fr^{2}}\hat{j}.\label{eq:7-1}
\end{equation}

\noindent Using the predicted velocity field $\mathbf{U^{*}}$, new
time level pressure field is obtained by solving a pressure Poisson
equation (obtained from Eq.~(\ref{eq:volume conservation}) using
a predictor-corrector method); given as

\begin{equation}
\nabla\cdot\left(\frac{\nabla P^{n+1}}{\chi_{i}^{n}}\right)=\frac{1}{\Delta\tau}\nabla\cdot\mathbf{U^{*}}.\label{eq:9-1}
\end{equation}

\noindent Finally, by subtracting Eq.~(\ref{eq:7-1}) from Eq.~(\ref{eq:6})
and neglecting the velocity correction in the diffusion terms, continuity
satisfying velocity field at the new time level is obtained as

\begin{equation}
\mathbf{U}_{P}^{n+1}=\mathbf{U}_{P}^{*}-\frac{\nabla P^{n+1}}{\chi_{i}^{n}}\Delta\tau.\label{eq:8}
\end{equation}

\noindent While solving the pressure Poisson equation (Eq.~(\ref{eq:9-1})),
an interfacial jump boundary condition for pressure is used; presented
in the next subsection.

\subsubsection{Implementation of Jump Boundary Condition for Pressure\label{sub:Implementation-of-Jump}}

In a two-fluid system, there will be a lower dimensional mass-less
interface separating different fluids. As shown in Fig.~\ref{fig:Computational-domain-with}($a$),
for each fluid, there will be two types of control volumes: (1) interfacial
and (2) interior. While solving pressure Poisson equation (Eq.~(\ref{eq:9-1}))
for interfacial control volumes, as depicted by the computational
stencil in Fig.~\ref{fig:Computational-domain-with}($b$)-($c$), the resulting
linear algebraic equation involves pressure from the other fluid that
leads to a poor approximation of pressure gradient term across the
interface. This was demonstrated by Liu et al. \cite{liu2000} using
order-of-magnitude analysis.

The poor approximation is avoided by using a pressure jump as an interfacial
boundary condition \cite{liu2000} while solving the pressure Poisson
equation (Eq.~(\ref{eq:9-1})).  The interfacial pressure boundary
condition is obtained by incorporating force balance at the interface \cite{shaikh2018};
given as  ${(p_{1}-p_{2})-\mathbf{((\hat{n}\cdot\sigma)_{1}-(\hat{n}\cdot\sigma)_{2})\cdot\hat{n}}=-\gamma\mathbf{\nabla\cdot\hat{n}}}$,
here, $p$ is pressure, $\mathbf{\sigma}$ is viscous stress tensor, $\gamma$ is
surface tension coefficient and $\mathbf{\hat{n}}$ is a normal unit vector;
and subscripts 1 and 2 denote fluid-1 and fluid-2, respectively. This
force balance at the interface  takes care of the discontinuity in
pressure across the two-fluid interface.  Here, a finite volume method
based generalized algebraic formulation of pressure Poisson equation
(along with the interfacial jump boundary condition for pressure),
proposed by Shaikh et al. \cite{shaikh2018}, is used. The generalized
formulation involves an additional source term for the interfacial
control volumes that is zero for the interior control volumes for
pressure.

\begin{center}

\par\end{center}

\subsubsection{Special Treatment for a Non-Uniform Grid}

\begin{figure}
\begin{centering}

\par\end{centering}

\begin{centering}
\includegraphics[width=14cm]{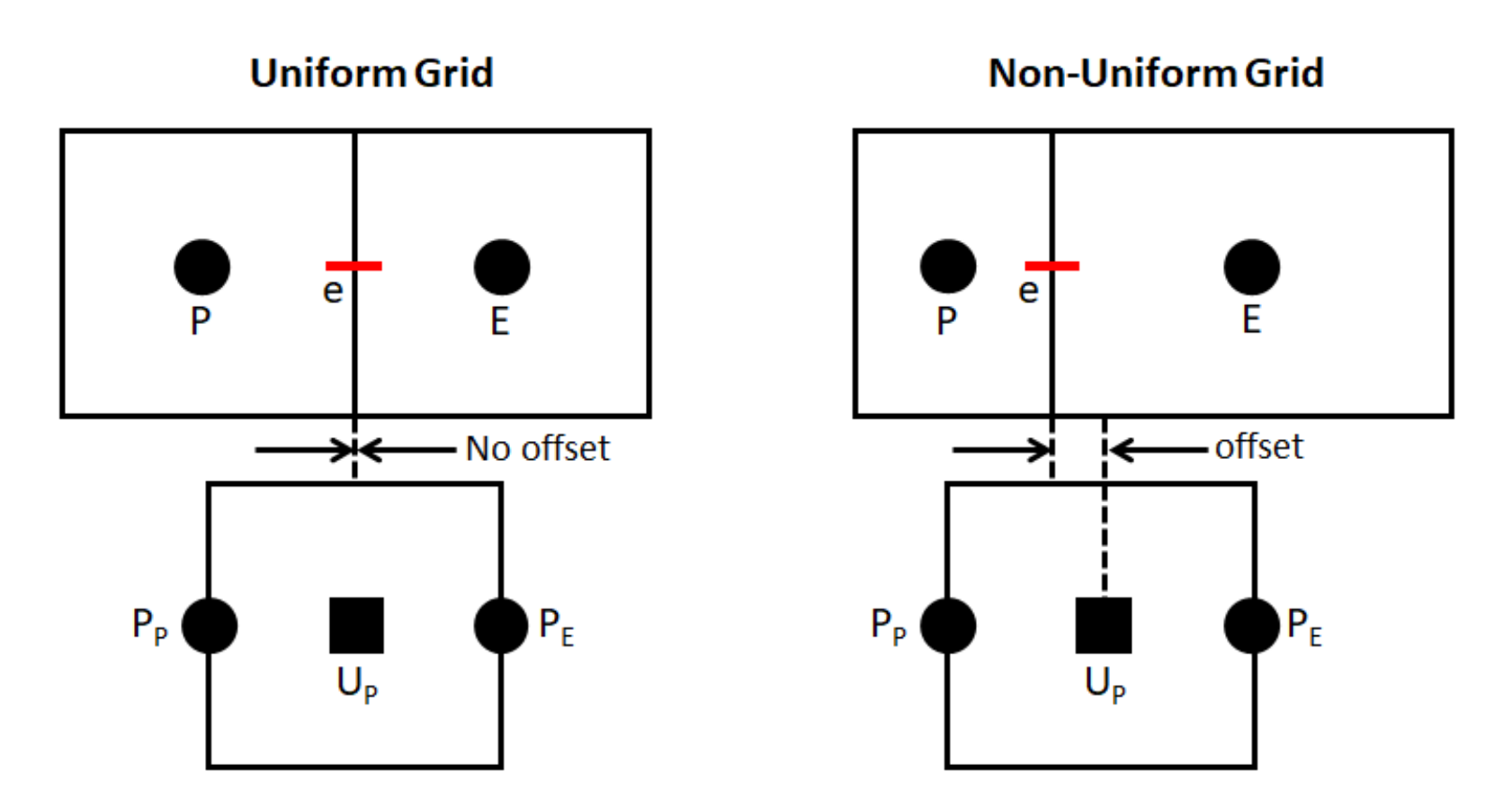}
\par\end{centering}

\protect\caption{\label{fig:Offset-between-velocity}Offset between the centroid of velocity
control volume and the face of pressure control volume.}

\end{figure}

Although Section \ref{sec:Concept-of-Semi-Adaptive} and Fig.~\ref{fig:(a)-Distribution-of}
presents AiMuR on a uniform grid, the AiMuR and SI-LSM based in-house
code is developed in the present work for both uniform and non-uniform
Cartesian-grid. In this section, additional numerical details for
non-uniform as compared to a uniform grid is presented. The essential
difference in the numerical methodology is due to the staggered grid
that results in a non-coinciding or an offset between the centroid
of a velocity control volume and the associated face-center of the
pressure control volume. This offset is shown in Fig.~\ref{fig:Offset-between-velocity}
for the east face of the non-uniform pressure control volume along
with no such offset for the uniform grid.

The offset for the non-uniform grid results in a distance-based linear
interpolations to compute the predicted velocities at the various
face centers ($U_{e}^{*}$, $U_{w}^{*}$, $V_{n}^{*}$, and $V_{s}^{*}$)
of the pressure control volume from the predicted velocities  (Eq.~(\ref{eq:7-1}))
at the centroid of the adjoining velocity control volumes ($U_{P}^{*}$,
$U_{W}^{*}$, $V_{P}^{*}$, and $V_{S}^{*}$); thereafter, the $U_{e}^{*}$,
$U_{w}^{*}$, $V_{n}^{*}$, and $V_{s}^{*}$ are used to calculate
the predicted mass source on the right-hand side of the pressure Poisson
equation (Eq. (\ref{eq:9-1})). Furthermore, after obtaining the correct
velocity field $\mathbf{U}_{P}^{n+1}$ from Eq.~(\ref{eq:8}), the
cell-center velocities are linearly interpolated to compute the velocity
at the face-center of the pressure control volumes ($U_{e}^{n+1}$,
$U_{w}^{n+1}$, $V_{n}^{n+1}$, and $V_{s}^{n+1}$). Finally, the
face-center velocity are  interpolated to obtain velocity  at the corners of
the pressure control volumes that is used to advect the level set
function field.

\subsection{Solution of Level Set Advection~(Mass-Conservation) Equations}\label{sub:Solution-of-Level}

\begin{figure}
\begin{centering}
\includegraphics[scale=0.65]{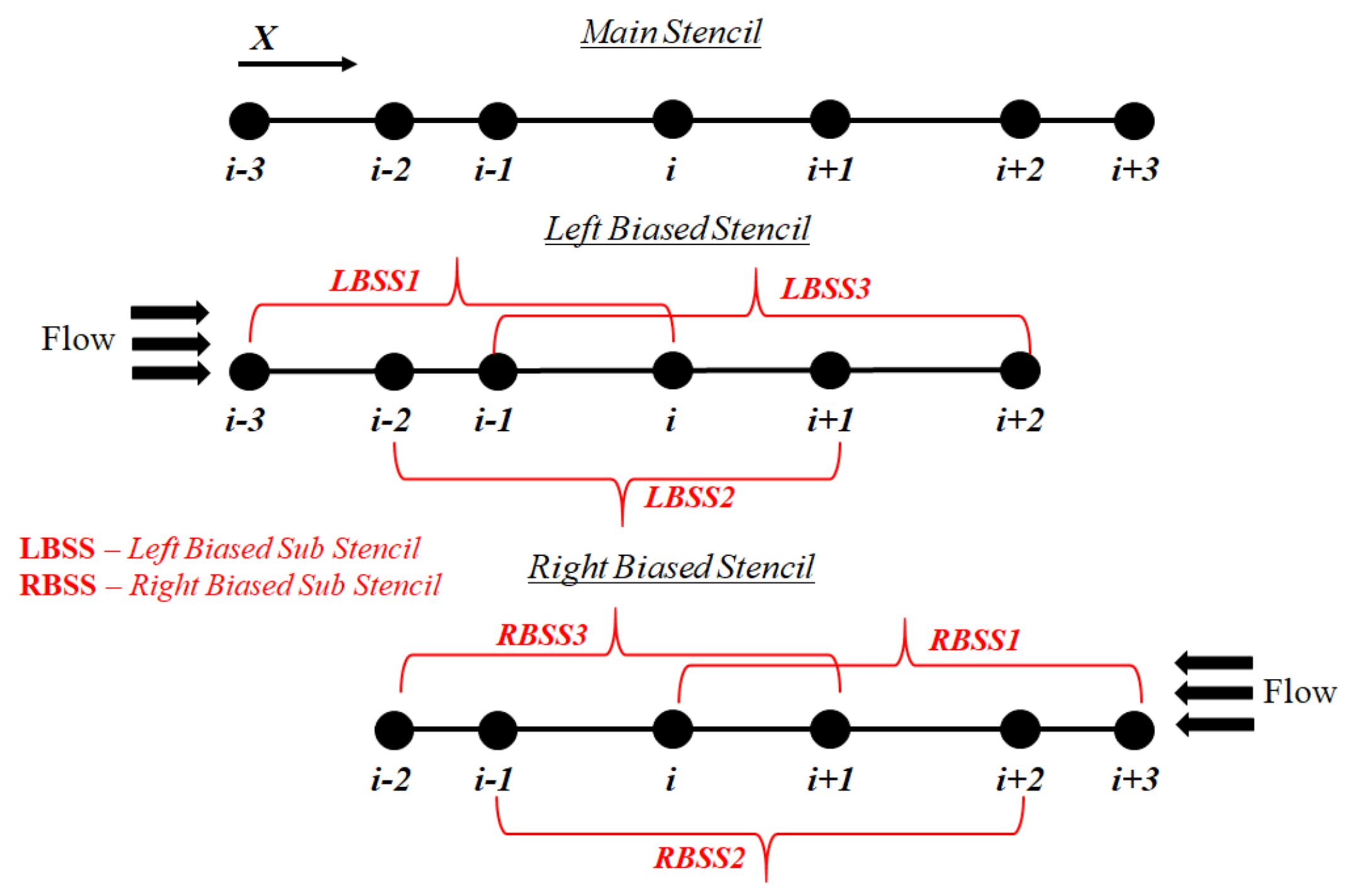}
\par\end{centering}

\protect\caption{\label{fig:Stencil-arrangement-in}Stencil arrangement for $3^{rd}$
order accurate Essentially Non-Oscillatory (ENO) scheme for non-uniformly
distributed grid points.}
\end{figure}

\begin{figure}
\begin{centering}
\includegraphics[scale=0.65]{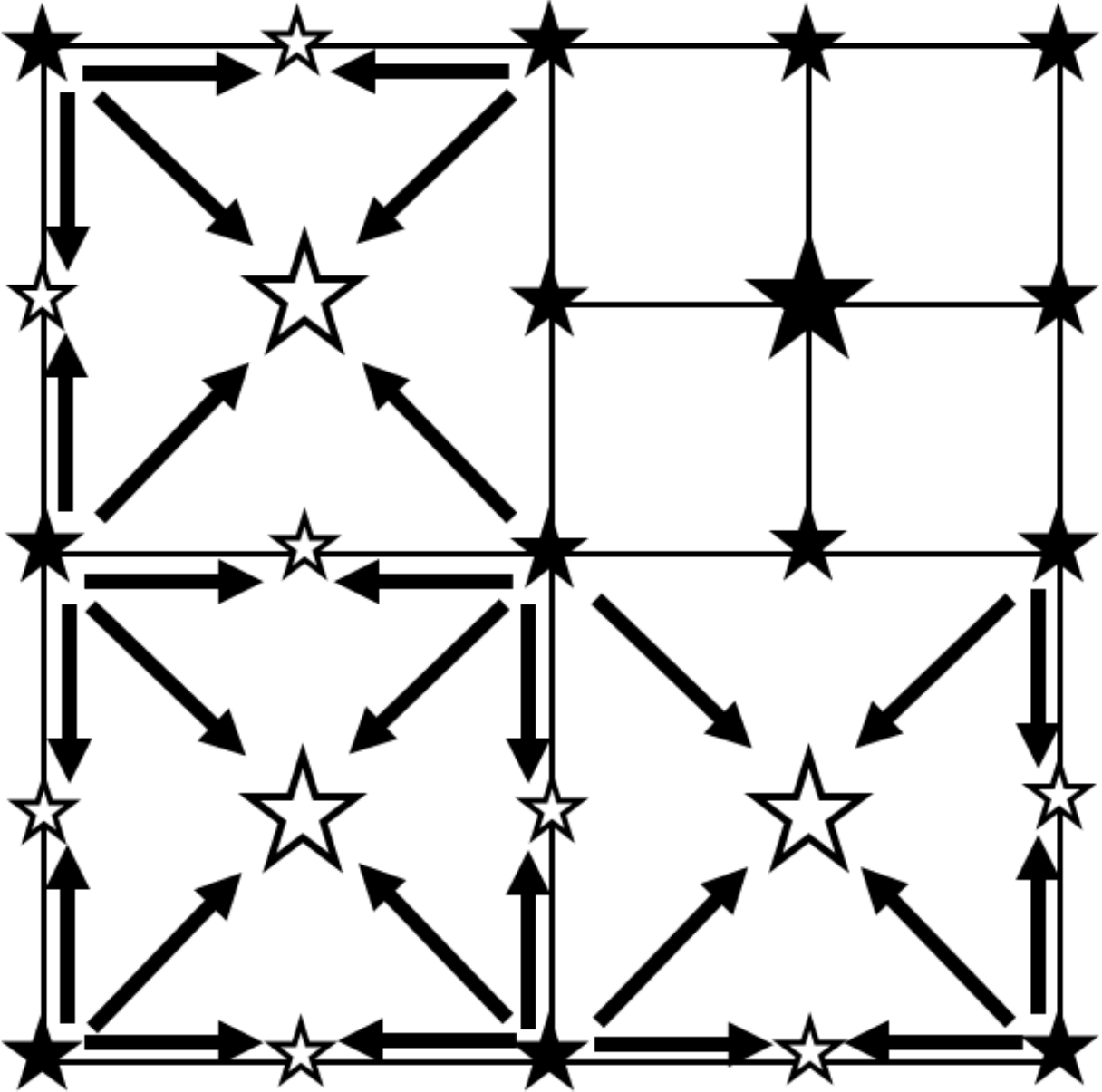}
\par\end{centering}

\protect\caption{\label{fig:Ghost-level-set} Interpolation of level set function at
ghost grid point using real level set function grid points.}
\end{figure}

The numerical methodology for the solution of the Navier-Stokes equations
(presented above) uses a physical law based finite volume method based algebraic formulation \cite{Sharma2017} while finite difference method
is used for the discretization of the level set equations (Eq.~(\ref{eq:lsadv})
and (\ref{eq:2.21})). Spatial (advection) term in the level set equations
is discretized using a III-order accurate Essentially Non-Oscillatory
(ENO) scheme (Jiang and Peng \cite{jiang2000}). The temporal discretization
of the level set advection equation is done using a III-order accurate
Runge-Kutta scheme and using a I-order accurate forward difference
for the reinitialization equation. Pseudo time step $\Delta\tau_{s}$
for the temporal term in the reinitialization equation is taken as
0.1 times of minimum grid size.

Although the formulation for the ENO scheme corresponding to a non-uniform
grid is available for the finite volume method in literature \cite{smit2005},
the formulation for finite difference method is presented here with
the help of Fig.~\ref{fig:Stencil-arrangement-in}. The figure shows
a non-uniform spacing based 7-point main-stencil, for implementing the ENO scheme on a non-uniform grid.
The figure also shows that the main stencil is subdivided into two
6-point stencil: left and right biased stencils (LBS and RBS). These
left/right stencils are further subdivided into three 4-point substencils:
left substencil as LBSS1, LBSS2, and LBSS3; and right substencil as
RBSS1, RBSS2, and RBSS3. Fitting a III-order Lagrange interpolation
polynomial in each of these substencils results in a finite difference
representation of the first derivative in the level set advection
equation $-$ presented below for $\phi_{X}$ (representing ${\partial\phi}/{\partial X}$)
at the various LBSS as

\begin{multline}
\phi_{X,i}^{LBSS1}=\phi_{i-3}\left(\frac{X_{i}^{2}-X_{i}X_{i-1}-X_{i}X_{i-2}+X_{i-1}X_{i-2}}{\left(X_{i-3}-X_{i}\right)\left(X_{i-3}-X_{i-2}\right)\left(X_{i-3}-X_{i-1}\right)}\right)\\
+\phi_{i-2}\left(\frac{X_{i}^{2}-X_{i}X_{i-1}-X_{i}X_{i-3}+X_{i-1}X_{i-3}}{\left(X_{i-2}-X_{i-3}\right)\left(X_{i-2}-X_{i-1}\right)\left(X_{i-2}-X_{i}\right)}\right)\\
+\phi_{i-1}\left(\frac{X_{i}^{2}-X_{i}X_{i-2}-X_{i}X_{i-3}+X_{i-2}X_{i-3}}{\left(X_{i-1}-X_{i-3}\right)\left(X_{i-1}-X_{i-2}\right)\left(X_{i-1}-X_{i}\right)}\right)\\
+\phi_{i}\left(\frac{1}{\left(X_{i}-X_{i-3}\right)}+\frac{1}{\left(X_{i}-X_{i-2}\right)}+\frac{1}{\left(X_{i}-X_{i-1}\right)}\right),\label{eq:2.22-2}
\end{multline}

\begin{multline}
\phi_{X,i}^{LBSS2}=\phi_{i-2}\left(\frac{X_{i}^{2}-X_{i}X_{i-1}-X_{i}X_{i+1}+X_{i-1}X_{i+1}}{\left(X_{i-2}-X_{i-1}\right)\left(X_{i-2}-X_{i}\right)\left(X_{i-2}-X_{i+1}\right)}\right)\\
+\phi_{i-1}\left(\frac{X_{i}^{2}-X_{i}X_{i-2}-X_{i}X_{i+1}+X_{i-2}X_{i+1}}{\left(X_{i-1}-X_{i-2}\right)\left(X_{i-1}-X_{i}\right)\left(X_{i-1}-X_{i+1}\right)}\right)\\
+\phi_{i+1}\left(\frac{X_{i}^{2}-X_{i}X_{i-1}-X_{i}X_{i-2}+X_{i-1}X_{i-2}}{\left(X_{i+1}-X_{i}\right)\left(X_{i+1}-X_{i-2}\right)\left(X_{i+1}-X_{i-1}\right)}\right)\\
+\phi_{i}\left(\frac{1}{\left(X_{i}-X_{i-2}\right)}+\frac{1}{\left(X_{i}-X_{i-1}\right)}+\frac{1}{\left(X_{i}-X_{i+1}\right)}\right),\label{eq:2.28-1}
\end{multline}

\begin{multline}
\phi_{X,i}^{LBSS3}=\phi_{i-1}\left(\frac{X_{i}^{2}-X_{i}X_{i+1}-X_{i}X_{i+2}+X_{i+1}X_{i+2}}{\left(X_{i-1}-X_{i}\right)\left(X_{i-1}-X_{i+1}\right)\left(X_{i-1}-X_{i+2}\right)}\right)\\
+\phi_{i+1}\left(\frac{X_{i}^{2}-X_{i}X_{i-1}-X_{i}X_{i+2}+X_{i-1}X_{i+2}}{\left(X_{i+1}-X_{i-1}\right)\left(X_{i+1}-X_{i}\right)\left(X_{i+1}-X_{i+2}\right)}\right)\\
+\phi_{i+2}\left(\frac{X_{i}^{2}-X_{i}X_{i-1}-X_{i}X_{i+1}+X_{i-1}X_{i+1}}{\left(X_{i+2}-X_{i}\right)\left(X_{i+2}-X_{i+1}\right)\left(X_{i+2}-X_{i-1}\right)}\right)\\
+\phi_{i}\left(\frac{1}{\left(X_{i}-X_{i-1}\right)}+\frac{1}{\left(X_{i}-X_{i+1}\right)}+\frac{1}{\left(X_{i}-X_{i+2}\right)}\right),\label{eq:2.29-1}
\end{multline}

From the above values, the $\phi_{X,i}^{LBSS}$ is chosen as

\begin{equation}
\phi_{X,i}^{LBSS}=\begin{cases}
\phi_{X,i}^{LBSS1} & \begin{array}{cc}
if\end{array}\mid B\mid<\mid C\mid\&\mid A-B\mid<\mid B-C\mid\\
\phi_{X,i}^{LBSS3} & \begin{array}{cc}
if\end{array}\mid B\mid>\mid C\mid\&\mid B-C\mid>\mid C-D\mid\\
\phi_{X,i}^{LBSS2} & \text{otherwise},
\end{cases}\label{eq:2.22-1-1}
\end{equation}

where

\[
A=\frac{\phi_{i-3}-\phi_{i-2}}{X_{i-2}-X_{i-3}}+\frac{\phi_{i-1}-\phi_{i-2}}{X_{i-1}-X_{i-2}},\,\,\,\, B=\frac{\phi_{i-2}-\phi_{i-1}}{X_{i-1}-X_{i-2}}+\frac{\phi_{i}-\phi_{i-1}}{X_{i}-X_{i-1}}
\]

\[
C=\frac{\phi_{i-1}-\phi_{i}}{X_{i}-X_{i-1}}+\frac{\phi_{i+1}-\phi_{i}}{X_{i+1}-X_{i}},\,\, \text{and}\,\, D=\frac{\phi_{i}-\phi_{i+1}}{X_{i+1}-X_{i}}+\frac{\phi_{i+2}-\phi_{i+1}}{X_{i+2}-X_{i+1}}.
\]

Similarly, the expression for $\phi_{X,i}^{RBSS}$ can be obtained
and the ENO scheme based discretized form of $\phi_{X,i}$ in the level
set advection equation (Eq.~(\ref{eq:lsadv})) is given as

\begin{equation}
U\frac{\partial\phi}{\partial X}=\text{max}(U,0)\phi_{X,i}^{LBSS}+\text{min}(U,0)\phi_{X,i}^{RBSS}
\end{equation}

For the AiMuR
on a uniform or non-uniform Cartesian-grid, now the implementation details for the above ENO scheme are discussed. \review{For the AiMuR, ghost grid points are considered in the stencil wherever needed while applying the ENO scheme for the real grid points; and are not needed for the interfacial boundary condition since we ensured a sufficiently wider band of the refined grid near the interface.}
 This is done to ensure that the weights of neighboring $\phi'$s in Eq.
(\ref{eq:2.22-2}), (\ref{eq:2.28-1}) and (\ref{eq:2.29-1}) are
computed only once (after the generation of uniform or non-uniform
Cartesian-grid) and do not change with time $-$ they are not dynamic.
Values of the level set function at the ghost grid points are computed
by linear interpolation of the adjoining real grid points. This
is shown in Fig.~\ref{fig:Ghost-level-set}, where the arrows show
the neighboring real child point values involved in the interpolation.
In addition to the computation of ENO scheme, these interpolated level
set function values at ghost grid points are utilized when ghost level
set grid points turn into real level set grid points.

\section{Solution Algorithm }
\begin{enumerate}
\item Generate initial configuration of fluid-fluid interface for all the
level set grid points. Initialize pressure and velocity as zero. Calculate
the weights of ENO scheme based on the distribution of level set grid
points.
\item Identify parent as well as child level set grid points (see section
\ref{sec:Concept-of-Semi-Adaptive}). Further, bifurcate them into
real and ghost level set grid points.
\item Calculate the thermo-physical properties using a sharp Heaviside function
\cite{shaikh2018}. Harmonic mean of the thermo-physical properties on
the either side of the interface is considered for interfacial cells.
\item Calculate the advection and diffusion flux (in Eq.~(\ref{eq:7-1}))
considering the continuity in velocity field and in its gradient at
the fluid-fluid interface. Predict the velocity $\mathbf{U^{*}}$
at the new time level by solving Eq.~(\ref{eq:7-1}). Here, III-order
Lin-Lin \textit{total variation diminishing} (TVD) \cite{date2005} scheme
is employed for discretizing the explicit advection term while the
diffusion term is discretized using a central difference scheme.
\item Calculate the mass-source (RHS of Eq.~(\ref{eq:9-1})) by linearly
interpolating predicted velocity $\mathbf{U^{*}}$ at the face-center
of the pressure control volume.
\item Obtain the converged solution of the pressure Poisson equation (Eq.~(\ref{eq:9-1})) using the jump condition by Ghost Fluid Method
(GFM).
\item Calculate the corrected velocity at the new time step (Eq.~(\ref{eq:8})).
Linearly interpolate the corrected velocity at the face-center of
the pressure control volume.
\item Obtain the level set advection velocity $\mathbf{U}_{a}$ at the real
level set grid points using a linear interpolation of the velocity
at the face-center of the pressure control volume. Advect the level
set function field using Eq.~(\ref{eq:lsadv}) for real level set
grid points using the methodology explained in subsection \ref{sub:Solution-of-Level}.
\item Interpolate the advected level set function field at ghost level set
grid points from the real level set grid points (see Fig.~\ref{fig:Ghost-level-set}).
\item Set the level set field as normal signed distance function by solving
level set reinitialization equation (Eq.~\ref{eq:2.21}) for real
level set grid points.
\item Repeat step 9.
\item Go to step 2 until the stopping criterion is met.
\end{enumerate}

\section{Validation and Qualitative Performance Study of A\lowercase{i}M\lowercase{u}R based SI-LSM }\label{sec:validation}

\begin{table}
\protect\caption{\label{tab:Resolution-of-different}Grid size for the five different grid
types: uniform coarse grid ($U_{c}$), non-uniform coarse grid ($NU_{c}$),
non-uniform coarse grid with AiMuR ($NU_{c}^{AiMuR}$), uniform fine grid with AiMuR ($U_{f}^{AiMuR}$) and uniform fine grid ($U_{f}$)
corresponding to the dam break, jet breakup and droplet coalescence problem . }

\begin{centering}

\par\end{centering}

\centering{}%
\begin{tabular}{cccc}
\hline
 & DB & JB & DC\tabularnewline
\hline
$U_{c}$, $NU_{c}$, and $NU_{c}^{AiMuR}$ & 100$\times$50 & 35$\times$200 & 100$\times$200\tabularnewline
$U_{f}$ and $U_{f}^{AiMuR}$ & 144$\times$80 & 50$\times$300 & 200$\times$400\tabularnewline
\hline
\end{tabular}
\end{table}

\begin{figure}
\begin{centering}
\includegraphics[width=16cm]{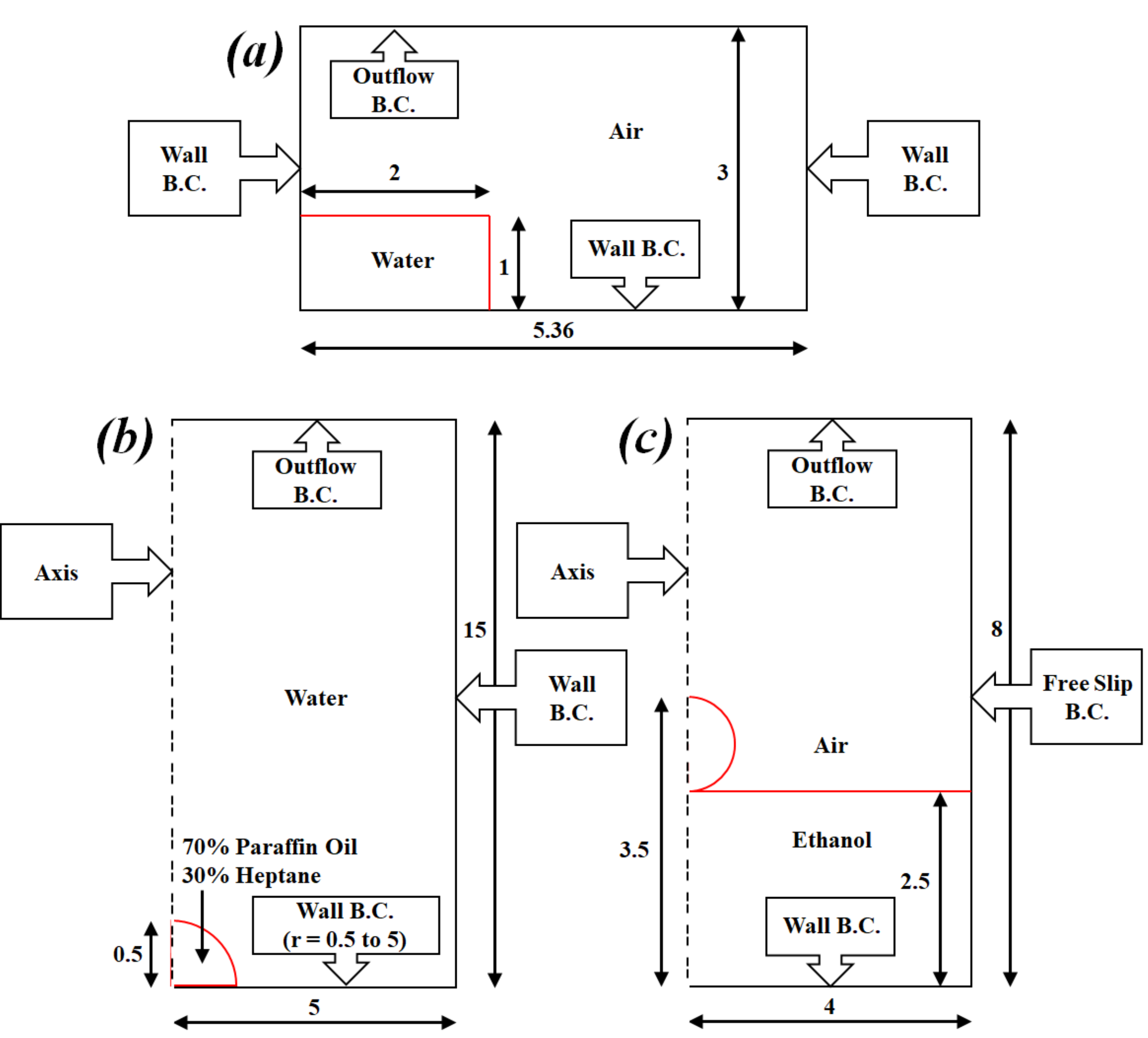}
\par\end{centering}

\protect\caption{\label{fig:Computational-setup-for-3}Computational setup for (a)
Dam break simulation (b) Breakup of a liquid jet and (c) Coalescence
of a stagnant ethanol drop at air-ethanol interface.}
\end{figure}

In order to present the validation of the proposed numerical methodology
and performance study, three different types of two-phase flow problems are considered: \textbf{\underline{D}}am-\textbf{\underline{B}}reak
(\textbf{DB}), \textbf{\underline{J}}et-\textbf{\underline{B}}reakup
(\textbf{JB}) and \textbf{\underline{D}}rop-\textbf{\underline{C}}oalescence
(\textbf{DC}). The dominant force is gravity, inertia and capillary force in the DB, JB and DC simulations, respectively. The DB simulation does not involve breakup of interface while the JB and DC problems involve more rigorous interface dynamics with break-up of interface that leads to a droplet formation. Computational setup corresponding to the three problems are shown in Fig.
\ref{fig:Computational-setup-for-3}. A performance study of the proposed AiMuR based SI-LSM is presented here by considering the adaptive unrefinement of the interface mesh on both uniform and non-uniform grid. However, since the result on a non-uniform as compared to the uniform grid is more accurate, the AiMuR on a non-uniform grid is considered on a coarser grid while the AiMuR on a uniform grid is presented on a finer grid; the respective AiMuR based SI-LSM is represented here as  $NU_{c}^{AiMuR}$ and $U_{f}^{AiMuR}$. Considering our in-house codes for the novel AiMuR based SI-LSM as well as the traditional SI-LSM, the scope of the present performance study is to compare the relative accuracy of the novel and traditional SI-LSMs (on uniform and non-uniform grid), with the accuracy obtained by comparing with the published experimental and numerical results for the DB, JB and DC problems. The relative accuracy is presented qualitatively in this section and and quantitatively in the next section. The resulting five different grid types of SI-LSM are presented in Table \ref{tab:Resolution-of-different} along with the associated grid size considered in the present simulations. Note that grid size mentioned in Table \ref{tab:Resolution-of-different} for $NU_{c}^{AiMuR}$ and $U_{f}^{AiMuR}$ is without performing unrefinement for the level set function.

For the five grid types  ($U_{c}$, $NU_{c}$, $NU_{c}^{AiMuR}$, $U_{f}^{AiMuR}$, and $U_{f}$), the grid resolution of uniform coarse grid $U_{c}$ (Table \ref{tab:Resolution-of-different}) is intentionally
chosen such that, numerical result will not be accurate enough while the finer grid size $U_{f}$ is kept fine enough to produce reliable numerical results. Furthermore, non-uniform
coarse grid ($NU_{c}$) is chosen such that it comprises of same number of control volumes
as that of uniform coarse grid ($U_{c}$). However, grid stretching
in $NU_{c}$ is done such that the grid resolution in interfacial region is comparable to uniform fine grid  ($U_{f}$).  For grid case $NU_{c}^{AiMuR}$, control
volume distribution for pressure and velocity is same as that for
$NU_{c}$. Nevertheless, mesh unrefinement strategy is incorporated
in $NU_{c}^{AiMuR}$, which creates level set resolution equivalent
to $NU_{c}$ in interfacial region and coarser resolution in non-interfacial
region. Similar discussion is also applicable for grid types $U_{f}$ and $U_{f}^{AiMuR}$. Similar to $NU_{c}$, both $NU_{c}^{AiMuR}$
and $U_{f}^{AiMuR}$ possess grid resolution equivalent to that of $U_{f}$ near the interface.

Based on the characteristic of the grid types selected in present
work, one can predict that computational time for $U_{f}$ will be
maximum among all, which will get reduced after applying mesh unrefinement
($U_{f}^{AiMuR}$). Computational time associated with $U_{c}$ and
$NU_{c}$ should be nearly same, as they have same number of control volumes.
However, it  largely depends on the trend of iterations while solving
pressure Poisson equation and also up to some extent on additional
computational operations required for $NU_{c}$ as compared to $U_{c}$. Computational time
for $NU_{c}^{AiMuR}$ should be  less than that required by $NU_{c}$.
Difference in computational time for $NU_{c}^{AiMuR}$
($U_{f}^{AiMuR}$) and $NU_{c}$ ($U_{f}$) will depend on the number
of unrefined level set grid points (which implicitly depends on how the interface evolves with time) in $NU_{c}^{AiMuR}$ ($U_{f}^{AiMuR}$). The hypothesised computational time for the various grid types are compared quantitatively in the next section.

\subsection{Dam Break Simulation}\label{sub:Dam-Break-Simulation}

\begin{figure}
\begin{centering}

\par\end{centering}

\begin{centering}
\includegraphics[width=16.5cm]{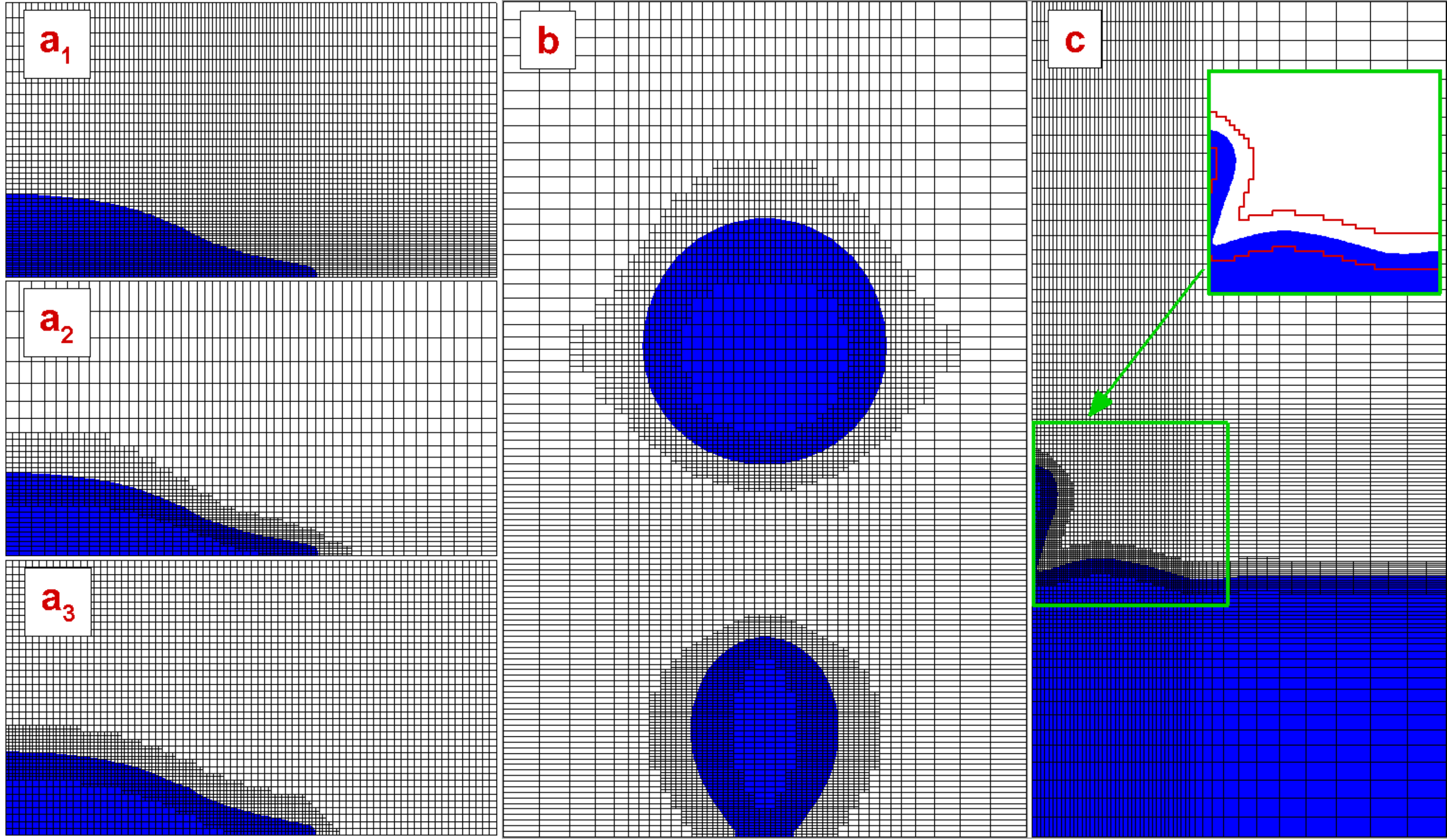}
\par\end{centering}

\protect\caption{\label{fig:Mesh-distribution-for} Interface (level set) mesh
for ($a$) dam break simulation, ($b$) liquid jet-breakup, and ($c$) drop coalescence
problems. The interface mesh shown in $(a_1)$, $(a_2,b,c)$, and $(a_3)$ is for $NU_{c}$, $NU_{c}^{AiMuR}$, and $U_{f}^{AiMuR}$, respectively. The \textit{adaptive unrefined instantaneous interface mesh} is at $\tau=1.5$ for ($a_2$) and ($a_3$), $\tau=150$ for ($b$), and $\tau=0.55$ for ($c$).}

\end{figure}

\begin{figure}
\begin{centering}

\par\end{centering}

\begin{centering}
\includegraphics[width=14.5cm]{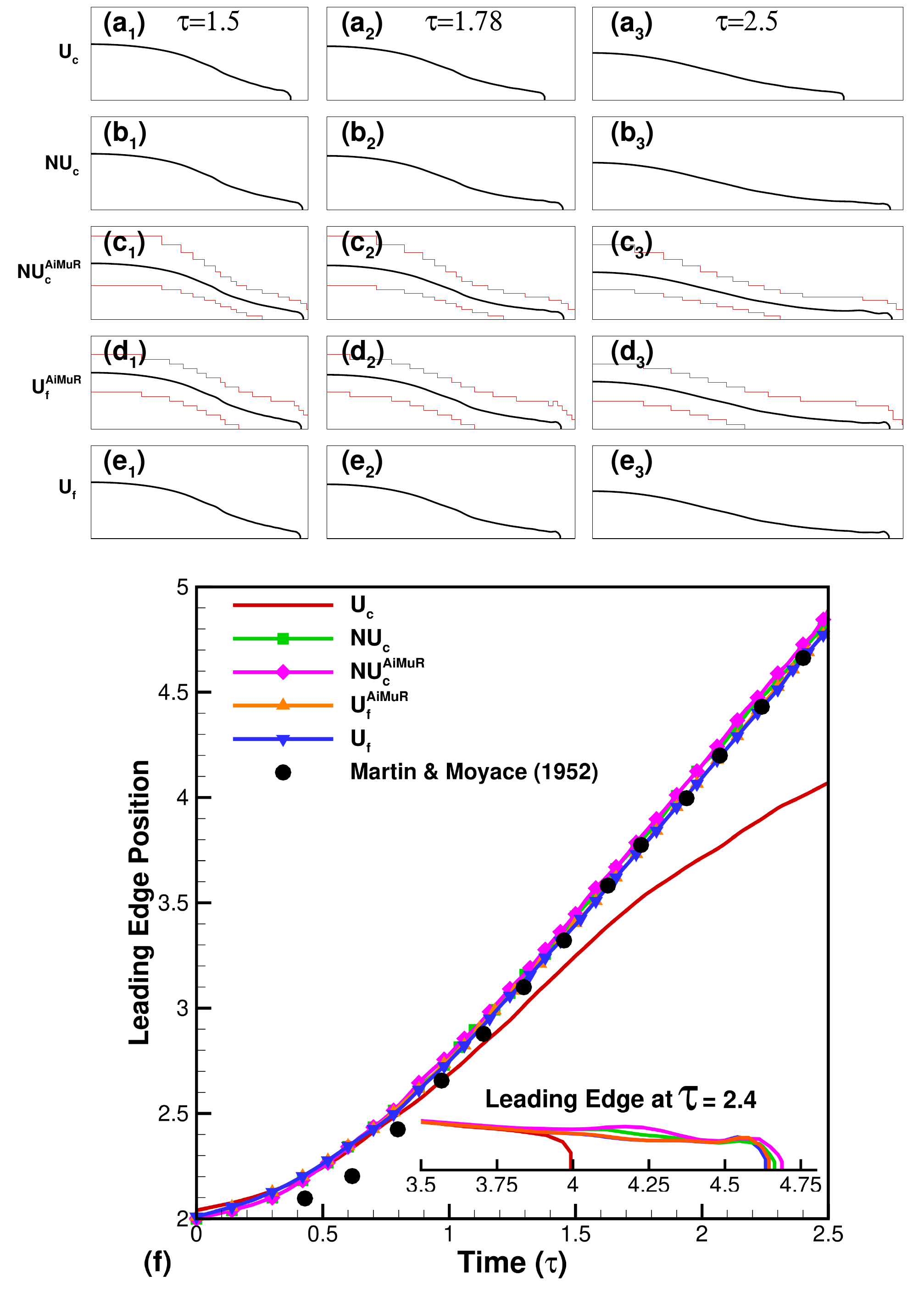}
\par\end{centering}




\protect\caption{\label{fig:Qualitative-and-quantitative}Dam break simulation at $Re=3 \times 10^6$ and $Fr=1$. For the present SI-LSM on the various grid types (Table \ref{tab:Resolution-of-different}), ($a-e$)~instantaneous interface position  at three different time instances and ($f$)~comparison of the present numerical and the published experimental \cite{martin1952} result. The stair-stepped lines for $NU_{c}^{AiMuR}$ and $U_{f}^{AiMuR}$ represent the interfacial region (in between the lines) outside which the level set grid points are dynamically unrefined.}
\end{figure}

Computational setup for this problem in Fig.~\ref{fig:Computational-setup-for-3}($a$)
shows a water column that is allowed to collapse under the effect of gravity. For the non-uniform Cartesian grid generation, a grid transformation function \cite{hoffmann2000}
is used that is given as
\begin{equation}
X/Y=L\lambda\left(1+\frac{\text{sinh}\left(\beta\left(\xi-A\right)\right)}{\text{sinh}\left(\beta A\right)}\right),\label{eq:4.2-1}
\end{equation}

where,

\[
A=\frac{\text{ln}\left(\frac{1+\lambda\left(e^{\beta}-1\right)}{1+\lambda\left(e^{-\beta}-1\right)}\right)}{2\beta}.
\]

\noindent The value of different tuning parameters for non-uniform distribution
in X and Y directions are $\beta_{X}=2.5$, $\beta_{Y}=3.1$, $\lambda_{X}=0.466$, and
$\lambda_{Y}=0.066$. Resulting non-uniform grid distribution is shown
in Fig.~\ref{fig:Mesh-distribution-for}($a_{1}$) for $NU_{c}$ and Fig.~\ref{fig:Mesh-distribution-for}($a_{2}$) for $NU_{c}^{AiMuR}$; and Fig.~\ref{fig:Mesh-distribution-for}($a_{3}$) shows adaptive unrefined uniform grid $U_{f}^{AiMuR}$. Note that the adaptive unrefined grid in Fig.~\ref{fig:Mesh-distribution-for}($a_{2},a_{3}$) corresponds to $\tau=1.5$. For instantaneous interface position, Fig.~\ref{fig:Qualitative-and-quantitative}($a$)-($e$) shows excellent agreement between the present results on the various types of SI-LSM on uniform/non-uniform grids (with or without interface mesh unrefinement) except the present result on $U_{c}$. This is also demonstrated for the leading edge distance in Fig.~\ref{fig:Qualitative-and-quantitative}($f$) by comparing with a benchmark experimental results \cite{martin1952}.
\noindent

\subsection{Breakup of a Liquid Jet}\label{sub:Breakup-of-a}

\begin{figure}
\begin{centering}

\par\end{centering}

\begin{centering}

\par\end{centering}

\begin{centering}
\includegraphics[width=12cm]{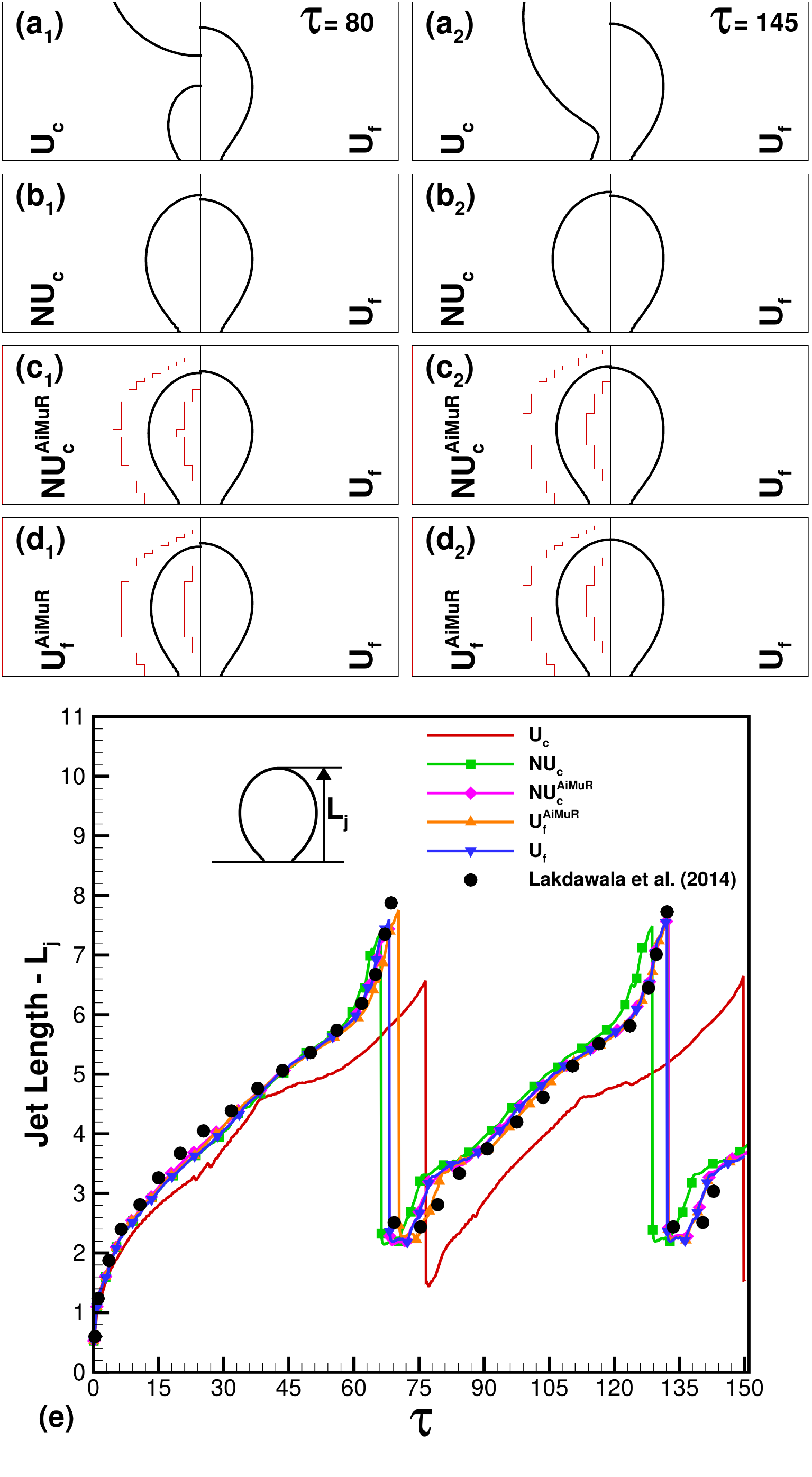}
\par\end{centering}


\protect\caption{\label{fig:Qualitative-and-quantitative-1}For jet breakup study at $Re=396$, $We=1.27$, and $Fr=5.44$, comparison of instantaneous interface obtained from $U_{c}$, $NU_{c}$, $NU_{c}^{AiMuR}$,
and $U_{f}^{AiMuR}$ (left) with that obtained from
$U_{f}$ (right) after $(a_{1}-d_{1})$ first jet breakup and $(a_{2}-d_{2})$ second jet breakup at time instances $\tau=80$ and $145$, respectively. Temporal variation
of jet length $L_j$ obtained in the present work is plotted in ($e$) and compared with the published numerical results \cite{lakdawala2014}.}
\end{figure}

Computational setup for this problem is shown in Fig.~\ref{fig:Computational-setup-for-3}($b$), where a lighter liquid is injected (against the gravity) in the heavier liquid with a constant velocity $0.15$  m/s. For the present problem, as long as the surface tension force dominates over the
buoyancy force, the jet will keep on rising. Once buoyancy force exceeds
the surface tension force, a neck forms and a droplet gets detached from the jet. The detached droplet continues to rise
in the heavier stagnant fluid and the jet will regain its original shape.

Number of control volumes employed for the present jet breakup problem is presented in Table \ref{tab:Resolution-of-different}. For
$NU_{c}$ and $NU_{c}^{AiMuR}$, gird clustering is implemented in both radial and axial
direction, that results in almost same grid resolution to that for the fine
uniform grid ($U_{f}$) in the breakup region. Distribution of level
set function grid points after performing unrefinement on a non-uniform
grid is shown in Fig.~\ref{fig:Mesh-distribution-for}($b$) at a time instance $\tau=150$. Results for the interface dynamics corresponding to breakup of two droplets from the inlet jet are shown in Fig.~\ref{fig:Qualitative-and-quantitative-1}. The instantaneous interface at $\tau=80$ and $\tau=145$ in Fig.~\ref{fig:Qualitative-and-quantitative-1}($a$)$-$($d$) shows an excellent agreement between the present results on a coarse non-uniform grid (with and without unrefinement) as compared to the uniform grid. Similar agreement between the results on the various grid types and also with the results reported in Lakdawala
et al. \cite{lakdawala2014} is shown in Fig.~\ref{fig:Qualitative-and-quantitative-1}($e$), for the temporal variation of jet length $L_{j}$; except for the result on uniform coarse grid $U_{c}$, which experiences late breakup of the jet as the interplay
between surface tension and buoyancy force is not captured well because
of insufficient grid cells in radial and axial direction.
\subsection{Coalescence of an Ethanol Droplet}\label{sub:Coalescence-of-an}

\begin{figure}
\begin{centering}
\includegraphics[width=16cm]{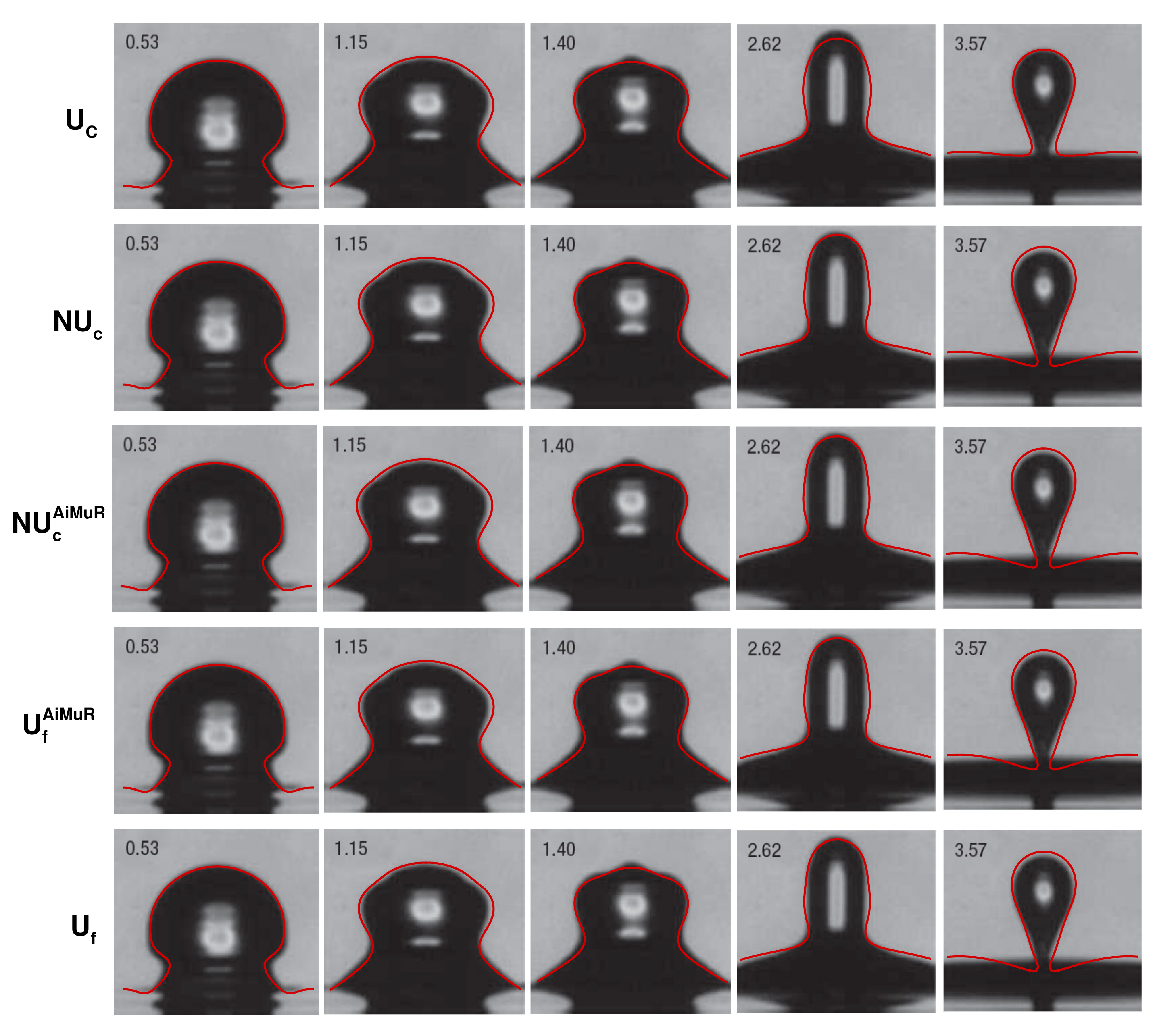}
\par\end{centering}





\protect\caption{\label{fig:Comparison-of-present}For coalescence study of an ethanol droplet (of diameter $1.07$ mm) in air, comparison of the temporal variation of present SI-LSM based numerically obtained instantaneous interface on five different grid types (Table \ref{tab:Resolution-of-different}) with experimental results of Blanchette and Bigioni \cite{blanchette2006}. The time instant marked above (0.53, 1.15, 1.40, 2.62 and 3.57) are in milli-second and the present numerically obtained interface is shown as line contour.}

\end{figure}

\begin{figure}

\begin{centering}
\includegraphics[width=14cm]{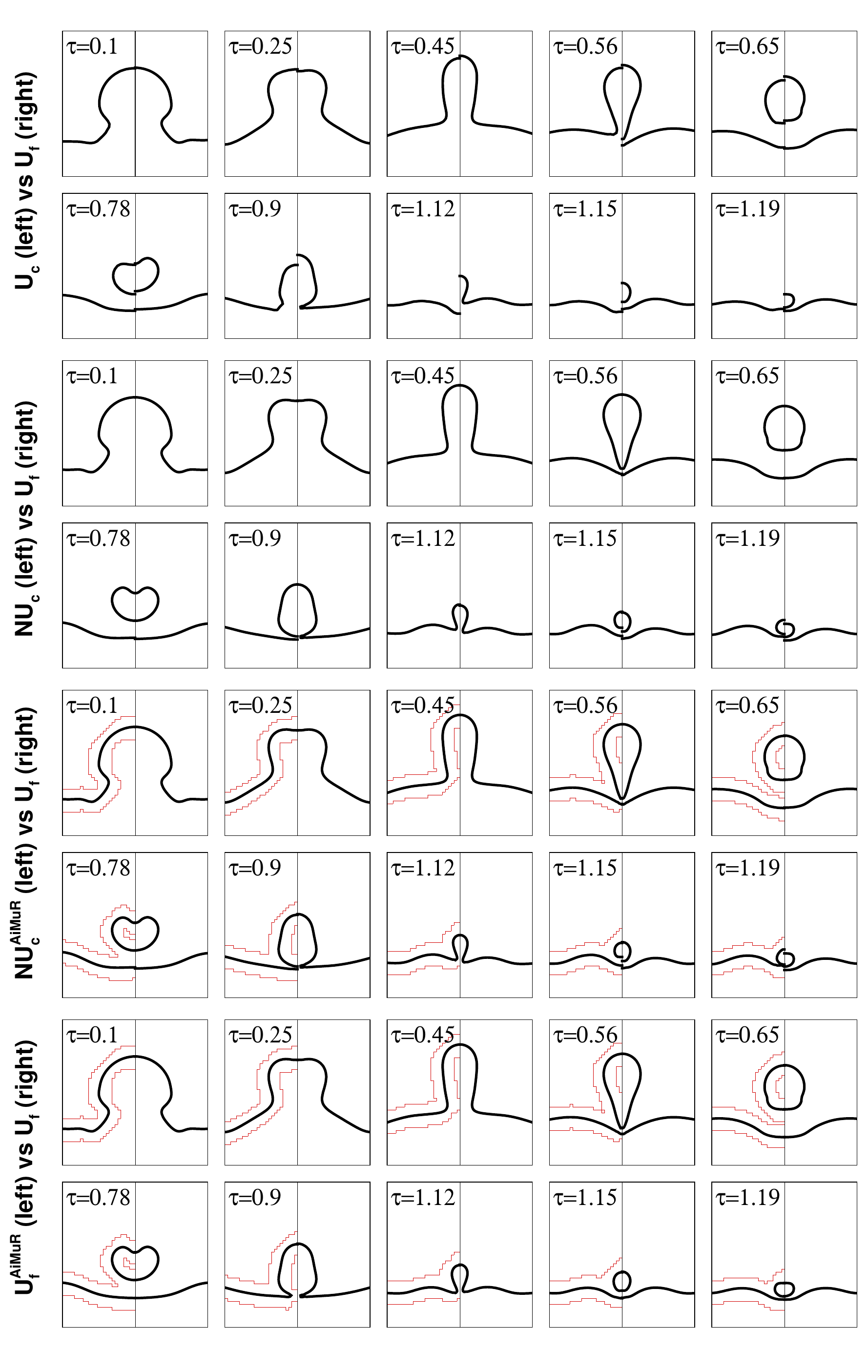}
\par\end{centering}








\protect\caption{\label{fig:Comparison-of-interface}
For coalescence dynamics of an ethanol droplet in air, comparison of temporal variation of instantaneous interface obtained for $U_{c}$, $NU_{c}$, $NU_{c}^{AiMuR}$, and $U_{f}^{AiMuR}$ (left) with that obtained for $U_{f}$ (right).}
\end{figure}

Computational set-up for the coalescence dynamics of an ethanol droplet of diameter $1.07$ mm, surrounded by air, over a pool of ethanol is shown in Fig.~\ref{fig:Computational-setup-for-3}($c$)); and the unrefined non-uniform interface-mesh at $\tau = 0.55$ is shown in Fig.~\ref{fig:Mesh-distribution-for}($c$). Fig.~\ref{fig:Comparison-of-present} shows an excellent agreement between the instantaneous interface obtained on the various grid types and the experimental results reported by Blanchette and Bigioni \cite{blanchette2006}. However, at time instance $3.57$ ms, the present result for $U_{c}$ as compared to the result on other grid types shows a much thicker neck that indicates a delay in the pinch-off and the resulting formation of secondary droplet.

As compared to the time-duration just before the first pinch-off, Fig.~\ref{fig:Comparison-of-interface} shows the temporal variation of the instantaneous interface for a longer time-duration corresponding to the second pinch-off of the secondary droplet. The figure shows that our result for the non-uniform grid with or without unrefinement agrees very well with the result on the finer uniform grid. However, the present result on a coarser uniform grid shows a slight delay in the first pinch-off of the primary droplet and thereafter it does not show the second pinch-off that is seen in the present results on the other grid types.

\review{

\section{Unrefinement Time-Period Independence and Order-of-Accuracy Studies}

\subsection{Unrefinement Time-Period Independence Study}\label{sub:Unref_freq}

\begin{figure}

\begin{centering}
\includegraphics[width=14cm]{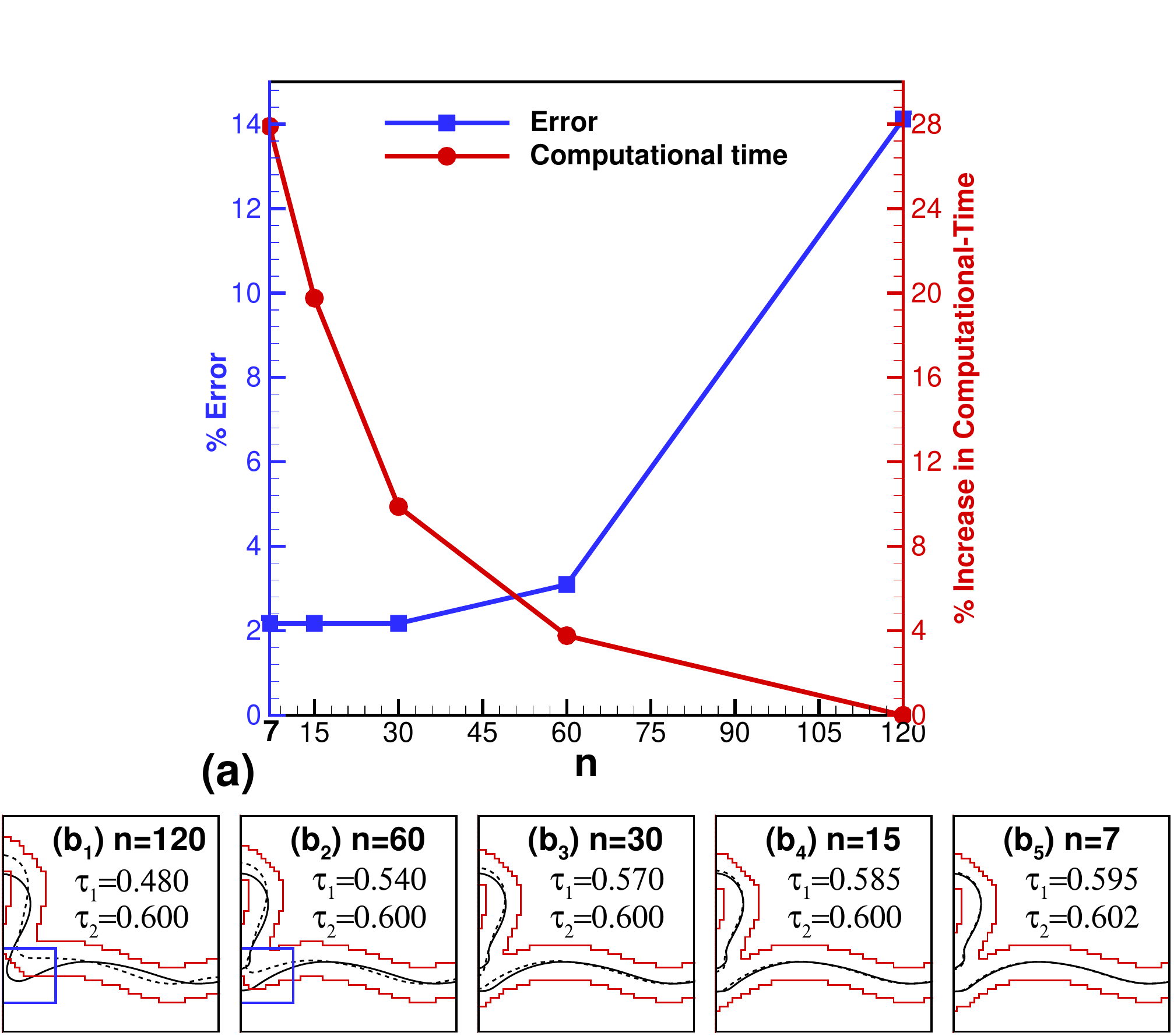}
\par\end{centering}



\protect\caption{\label{fig:adaption_frequency1}
\review{For the droplet coalescence problem, unrefinement time-period independence study of the present AiMuR on a uniform $200 \times 400$ grid: ($a$) variation of \% error and \% increase in computational-time with decreasing number ($n$)  of the time-step $\Delta t$ after which the interface-mesh is periodically-unrefined; and ($b_{1}$)-($b_{5}$) instantaneous interface obtained for the different values of $n$. The \% error is for the first pinch-off time reported by Blanchette and Bigioni \cite{blanchette2006}. For ($b_{1}$)-($b_{5}$), the region in-between the red-curves corresponds to the fine-mesh region, with unrefined coarser-mesh outside this region; and the dashed and solid black-lines represent the interface at a time-instant corresponding to the beginning ($\tau_{1}$) and end ($\tau_{2}$) of the associated unrefinement time-period, respectively.}} 

\end{figure}

For the present AiMuR, the periodic unrefinement of the interface mesh is done after certain number $n$  of the time-step $\Delta t$  that results in the time-period for the unrefinement as $n\Delta t$ . Thus  $n$  is a numerical parameter for the present method that is problem dependent and determined here from an unrefinement time-period independence study; similar to the grid-independence and time-step independence studies commonly used in CFD. The time-period independence study is presented in Fig.~\ref{fig:adaption_frequency1}($a$), for the uniform grid based AiMuR and the droplet-coalascence problem. The figure shows an asymtotic decrease in the error with increasing $n$, with almost no change in the error after $n=15$. Thus, $n=15$, \emph{i.e.}, the periodic-unrefinement after every $15^{th}$  time-step, is chosen for the AiMuR based simulation of the coalascence of an ethanol droplet in air. The figure also shows an almost $20\%$  increase in the computational time as $n$  decreases from $120$  to $15$.

The unrefinement after certain number of time-steps, instead of after every time-step, is due to the fact that the present AiMuR involves a wider band of finer mesh near the interface. The fine-mesh region, along with the interface extreme positions, during the time-interval of the unrefinement is shown in Fig.~\ref{fig:adaption_frequency1}($b_{1}$)-($b_{5}$) for various values of $n$. For the cases with $n=120$ and $60$, the figure shows that the interface moves outside the fine-mesh region during the time period $n\Delta t$  of the unrefinement; thus resulting in the larger error as seen in Fig.~\ref{fig:adaption_frequency1}($a$). The unrefinement time-period independence study ensures that the fluid-fluid interface stays within the fine-mesh region for an accurate solution. Similar unrefinement independence study for the other problems resulted in $n=50$ for the dam break problem and $n=150$ for the liquid jet breakup problem.

\subsection{Order of Accuracy Study}\label{sub:Order}

For a multiphase flow solver, order of accuracy study is widely reported for a problem on decaying oscillations of a capillary wave that has an analytical solution (Prosperetti \cite{prosperetti1981}). Thus, after ensuring the verification of our numerical with the analytical solution, Fig.~\ref{fig:order_of_accuracy1}($a$) presents an order of accuracy study for this problem. The computational setup for this problem can be found in Gerlach et al. \cite{Gerlach2006}; error is defined as $L_{2}$ norm of the difference between time-wise variation of the non-dimensional relative amplitude computed numerically and that obtained analytically. The figure shows that the order of accuracy of our SI-LSM on a uniform grid, with and without AiMuR, is between first and second order; almost same as that seen in the figure for the refined level-set grid method of Hermann \cite{hermann2008}. For a comparative study, the figure also presents the order of accuracy study of seven other numerical methods: front-tracking method \cite{ftm1999}, Gerris flow-solver \cite{popinet2009} (volume-of-fluid implementation with generalised height-function curvature calculation for quadtree and octree discretizations), PROST \cite{Renardy2002} (volume-of-fluid implementation with a parabolic reconstruction of surface tension) and CLSVOF \cite{Sussman2000} (coupled level-set volume-of-fluid formulation) implementations of Gerlach et al.\cite{Gerlach2006}, CSF-VOF \cite{Gueyffier1999}  (continuous surface force method based modelling of surface tension in volume-of-fluid framework), and VOF-NIFPA-1 \cite{Ivey2017} (VOF implementation with non-intersecting flux polyhedron advection (NIFPA) scheme for the advection of volume fraction) and Conservative DI \cite{Mirjalili2020} (diffuse-interface) implementations of Mirjalili et al. \cite{Mirjalili2019} . The figure shows that the Gerris, PROST, and CLSVOF are almost second order accurate while the other six multiphase flow solvers (including our SI-LSM) exhibit an order of accuracy between first and second order. It is worth noting from the figure that the present SI-LSM as compared to the other numerical methods is most accurate for the coarser grid resolution of $N=8$ Whereas, on relatively finer grid resolution of $N=64$, it can be seen that our SI-LSM is more accuracte than the front-tracking, CSF-VOF, conservative DI and RLSG.

Order of accuracy study is also presented in Fig.~\ref{fig:order_of_accuracy1}($b$) for the same dam-break simulation (Fig.~\ref{fig:Computational-setup-for-3}($a$)),considering the mass-error since this is the biggest disadvantage of the level set method \cite{shrma2015}. With grid refinement, the convergence of the mass-error (at some particular time-instant) in the figure shows that the present present AiMuR based LSM is somewhere between first and second order accurate; same as concluded from Fig.~\ref{fig:order_of_accuracy1}($a$). Using the physical interpretation of Heaviside function, proposed \cite{gada2009} and later used in the previous work from our reasearch group \cite{gada2011, Lakdawala2015}, the mass error is computed here as 

\begin{equation}
 \text{\% Mass Error} = 100\times\frac{1}{\tau_{max}}\int_{0}^{\tau_{max}}\frac{\left | \sum_{i,j}H_{i,j}^{\tau}\Delta V_{i,j}-\sum_{i,j}H_{i,j}^{o}\Delta V_{i,j} \right |}{\sum_{i,j}H_{i,j}^{o}\Delta V_{i,j}}d\tau, 
 \end{equation}
 
\noindent where, $H_{i,j}^{o}$ is the initial Heaviside function and $H_{i,j}^{\tau}$ is Heaviside function at a particular time instance $\tau$.

\begin{figure}

\begin{centering}
\includegraphics[width=16cm]{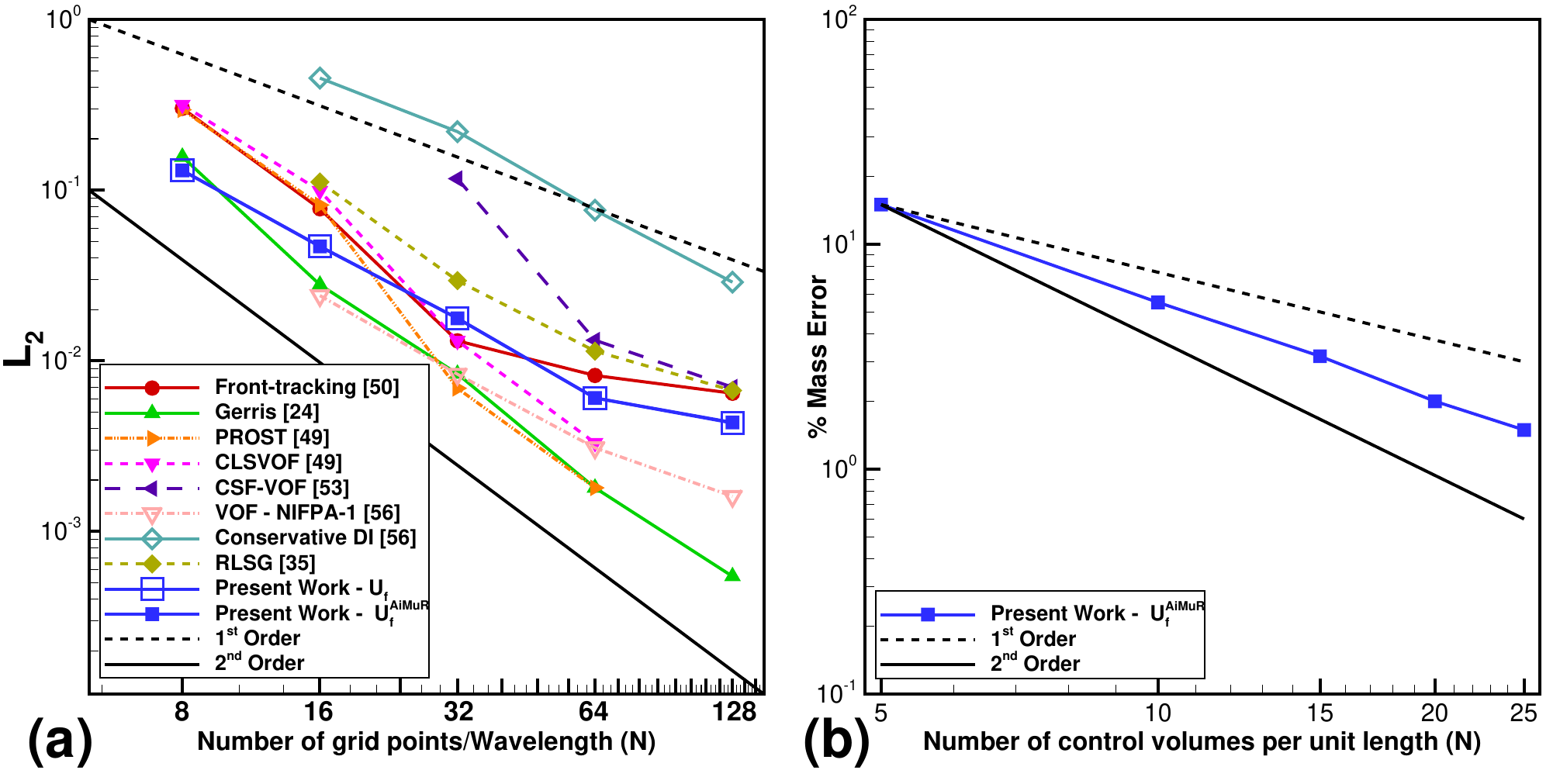}
\par\end{centering}

\protect\caption{\label{fig:order_of_accuracy1}\review{
With increasing grid refinement levels, variation of ($a$) $L_{2}$ norm of the error (between the present/published mumtiphase solvers and analytical solution \cite{prosperetti1981}) for the time-wise variation of the non-dimensional relative amplitude of a damped capillary wave (for $\rho_{1}=\rho_{2}=1$, $\mu_{1}=\mu_{2}=0.064720863$, and Ohnesorge number $Oh=1/\sqrt{3000}$) and (b) mass-error for the $U_{f}^{AiMuR}$  based dam-break simulation.}}

\end{figure}

}

\section{Quantitative Performance Study}\label{sec:performance}

\begin{figure}
\begin{centering}

\par\end{centering}

\begin{centering}
\includegraphics[width=16.5cm]{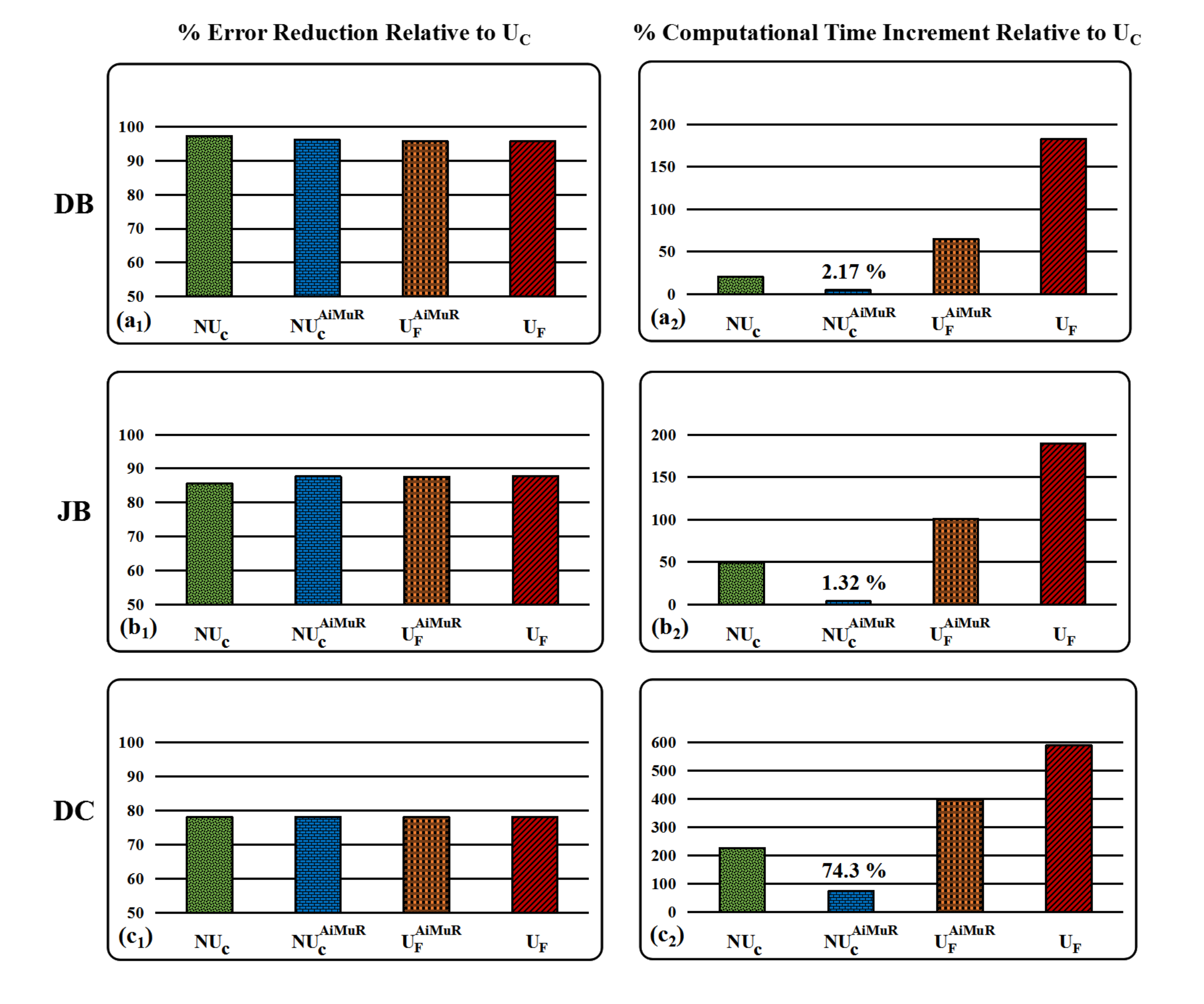}
\par\end{centering}

\protect\caption{\label{fig:Performance-curve-for}Variation of the performance parameters (Eq.~(\ref{eq:4.2-222})) for the present novel SI-LSM and traditional SI-LSM on the various grid types (Table \ref{tab:Resolution-of-different}) for $(a)$ dam break, $(b)$ jet breakup, and $(c)$ drop coalescence
problems.} 
\end{figure}

\begin{figure}
\begin{centering}
\includegraphics[width=16.5cm]{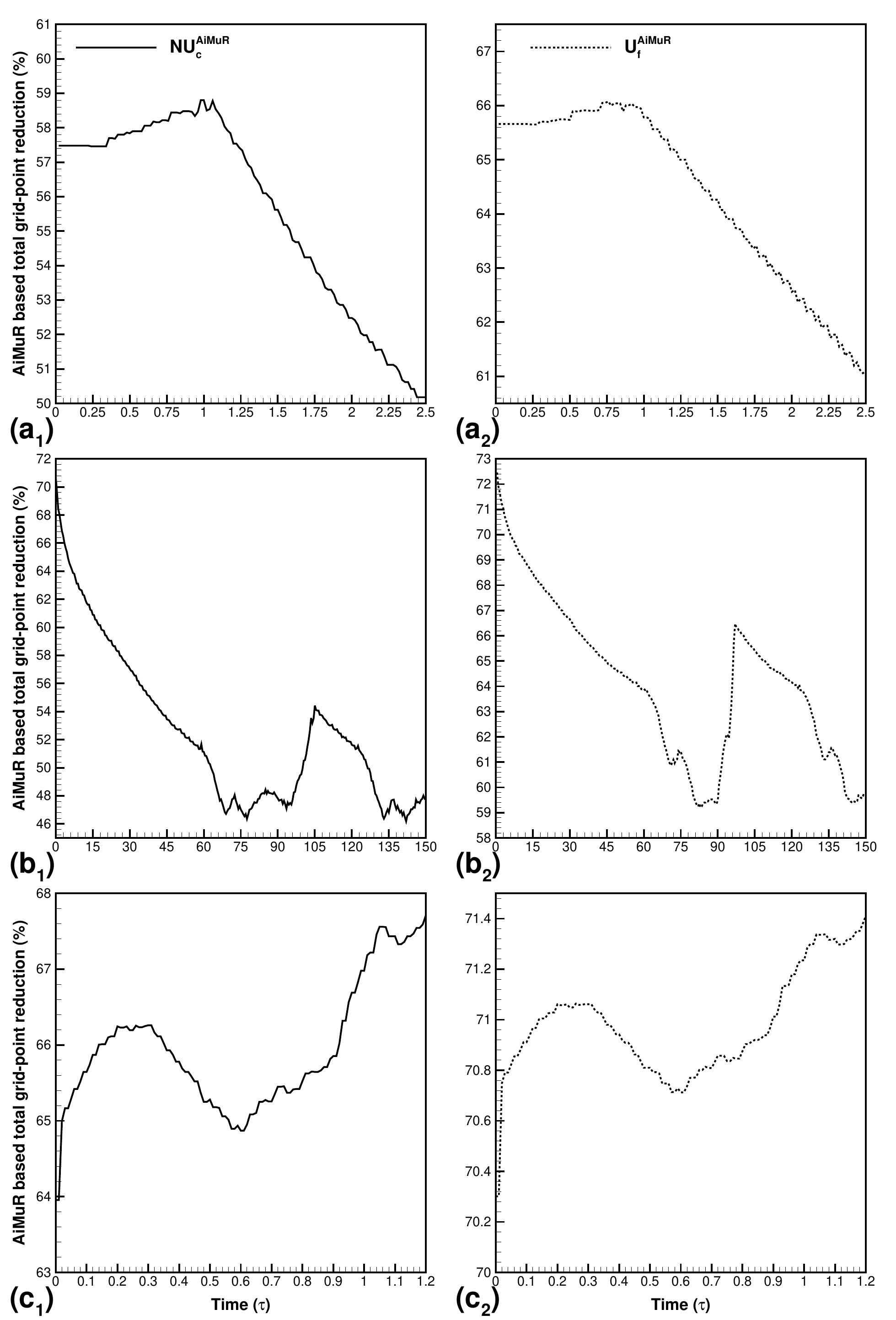}

\par\end{centering}



\protect\caption{\label{fig:Temporal-variation-of-1}\review{Temporal variation of AiMuR based
total grid-point reduction (\%) for the level set function on the $(a_1,b_1,c_1)$ non-uniform coarse grid $NU_{c}$ and $(a_2,b_2,c_2)$ uniform fine grid $U_{f}$, for $(a_1,a_2)$ dam break, $(b_1,b_2)$ jet breakup and $(c_1,c_2)$ droplet coalescence problems.}}
\end{figure}

For the three different two-phase problems, the above comparison of the unsteady interface dynamics on the various grid types and also with the published numerical/experimental results clearly demonstrates the superiority of the SI-LSM on the non-uniform grid and unrefinement as compared to that on the uniform grid. Almost similar superiority of the $NU_{c}$, $NU_{c}^{AiMuR}$,
and $U_{f}^{AiMuR}$ as compared to $U_{c}$ was demonstrated qualitatively above and presented quantitatively here.

The quantitative representation of the relative performance considers the \review{total} computational time \review{(including the time required for the interpolation and the inter-grid transfer)} along with the computational accuracy to distinguish the relatively superior performance of $NU_{c}$, $NU_{c}^{AiMuR}$,
and $U_{f}^{AiMuR}$ as compared to $U_{c}$. Thus,
for a quantitative representation, a detailed performance study is carried out by defining two performance parameters: \% Error
Reduction and \% Computational Time Increment; given as \review{ \cite{gada2011,Lakdawala2015,Patil2016,shaikh2019}}

\[
\%\,\, \text{Error}\,\, \text{Reduction}=\left(1-\frac{\text{Error}_{grid-type}}{\text{Error}_{U_{C}}}\right)\times100,
\]

\begin{equation}
\%\,\, \text{Computational}\,\, \text{Time}\,(C.T.)\,\, \text{Increment}=\left(\frac{C.T.{}_{grid-type}}{C.T.{}_{U_{C}}}-1\right)\times100,
\label{eq:4.2-222}
\end{equation}

where a grid-type corresponds to $NU_{c}$, $NU_{c}^{AiMuR}$,
$U_{f}^{AiMuR}$, and $U_{f}$. Further, present performance study is carried
out using coarse uniform grid ($U_{c}$) as a reference grid for all
the problems. In order to calculate the error associated with the present numerical
results on various grid types, as compared to the published experimental/numerical results, certain parameter is selected
in each problem.
The parameter considered for calculating the error of the present SI-LSM corresponds to leading edge distance reported by
Martin and Moyce \cite{martin1952} at time $\tau=2.4$, jet breakup length reported by Lakdawala et al. \cite{lakdawala2014} (after second
breakup) at time $\tau=145$, and first pinch-off time reported by Blanchette
and Bigioni \cite{blanchette2006} for\textbf{ }\textbf{\underline{D}}am-\textbf{\underline{B}}reak (\textbf{DB}),
\textbf{\underline{J}}et-\textbf{\underline{B}}reakup (\textbf{JB})
and \textbf{\underline{D}}rop-\textbf{\underline{C}}oalescence (\textbf{DC})
problems, respectively. 

For the three sufficiently different two phase flow problems, the performance parameters (Eq.~(\ref{eq:4.2-222})) for the present novel and traditional SI-LSM on various grid types are shown in Fig.~\ref{fig:Performance-curve-for}.
As discussed in previous subsection, the error reduction in Fig.~\ref{fig:Performance-curve-for} demonstrates a quantitative (in terms of accuracy) evidence of almost same superiority of the $NU_{c}$, $NU_{c}^{AiMuR}$,
and $U_{f}^{AiMuR}$ as compared to $U_{c}$. Whereas, the computational time increment clearly demonstrates the relative superiority of the $NU_{c}$, $NU_{c}^{AiMuR}$,
and $U_{f}^{AiMuR}$, with the least computational time increment by the SI-LSM on $NU_{c}^{AiMuR}$ as compared to that on $NU_{c}$ and $U_{f}^{AiMuR}$. Theoretically,
computational time for grid type $NU_{c}$ should be same as that
for $U_{c}$ (as both of them are having same number of control volumes).
Nevertheless, since $NU_{c}$ as compared to $U_{c}$ requires more iterations
for solving pressure Poisson equation, $NU_{c}$ results in more computational
time. Application of AiMuR on $NU_{c}$ reduces this overhead
in computational time. This is seen in Fig.~\ref{fig:Performance-curve-for}($a_{2},b_{2},$), with a negligible increase in computational time by $NU_{c}^{AiMuR}$ as compared to $U_{c}$. Further, the figure also shows that computational time demanded by $U_{f}$ gets reduced after applying the mesh unrefinement.

After applying AiMuR strategy
on either uniform or non-uniform grid, the reduction in computational time is correlated with AiMuR based total grid-point
reduction (\%) that is shown in Fig.~\ref{fig:Temporal-variation-of-1}. The figure shows that instantaneous value of AiMuR based total grid-point
reduction (\%) for level set function is more than 45\% for the problems studied here. Time-wise increase or decrease of AiMuR based total grid-point reduction
(\%) in Fig.~\ref{fig:Temporal-variation-of-1} also implicitly represents
the dynamics of fluid-fluid interface, \emph{i.e.}, spreading of the fluid-fluid interface in computational domain and AiMuR based total grid-point reduction (\%) are inversely related. In dam break problem, gradual
decrease in AiMuR based total grid-point reduction (\%) is due to
the interface spreading after the collapse of the water column. Similarly,
in jet breakup, initial decrement in total grid-point reduction (\%)
is because of the continuous injection of fluid. After the first breakup,
an increase in total grid-point reduction (\%) is attained as soon as the detached droplet escapes the computational domain, followed by another
decrease-increase cycle. For drop coalescence problem, an increase in
AiMuR based total grid-point reduction (\%) after first pinch off
corresponds to the smaller daughter droplet size. Among all grid combinations,
$NU_{c}^{AiMuR}$ is found to be computationally most efficient since it produces
numerical results of almost same accuracy as that on a fine uniform grid ($U_{f}$)
and requires a computational time almost same as that on coarse uniform grid
($U_{c}$).

\section*{Concluding Remarks}

In the present work, numerical methodology for simulating multi-phase flows on dynamically
unrefined uniform as well as non-uniform level set mesh is proposed,
where unrefinement is carried out away from the interface location.
The dynamic unrefinement is done for the Cartesian interface mesh corresponding to level set function only. Consequently,
higher order schemes based solution of level set equations (advection
and reinitialization) is obtained on almost half of the grid in the
region away from the interface. Further, ENO scheme
with varying weights is used to solve mass conservation equation on
highly stretched non-uniform grids. To demonstrate the numerical accuracy and
computational efficiency of the proposed AiMuR based SI-LSM, performance study
is carried out for three sufficiently different two-phase flow problems: dam break, breakup of a liquid jet and drop coalescence
problems. For a detailed qualitative and quantitative performance study of the proposed adaptive unrefinement based SI-LSM as compared to the traditional SI-LSM, a systematic combinations of various types of SI-LSM and coarse/fine grid size are considered. It is found that the present SI-LSM on a non-uniform coarse grid ($NU_{c}$) demands more computational
time than for the uniform coarse grid ($U_{c}$) with numerical accuracy
almost same as the uniform fine grid ($U_{f}$). After implementing
adaptive interface-mesh unrefinement (AiMuR) on a non-uniform grid ($NU_{c}^{AiMuR}$),
further reduction in computational time is obtained without much compromise
in numerical accuracy. Incorporating AiMuR on a fine uniform grid ($U_{f}^{AiMuR}$)
also produces results of almost same accuracy as that of uniform fine grid but with
less computational time. However, reduction in computational time
by $U_{f}^{AiMuR}$ is not as significant as that of $NU_{c}^{AiMuR}$.

\review{The application of the AMR as well as the AiMuR algorithm generates a time-evolving hierarchical distribution of Cartesian control volumes. The evolution of this hierarchical distribution (successive refinement or un-refinement of control volumes) is governed by the mesh refinement/un-refinement criteria that is based on flow physics as well as interface dynamics for the AMR and only interface dynamics for the AiMuR. However, the implementation details of the novel AiMuR are relatively less complex as compared to the AMR. The AiMuR based LSM is presented here as a proof-of-concept and studies on the performance of AiMuR on more suitable multiphase problems is part of future work.} Application of the present non-uniform and adaptive unrefinement grid strategies will be extended
to two-phase flow involving phase change. Furthermore, performance characteristics of these grid types
will also be studied for three-dimensional multi-processor multi-phase flow simulations as present study is restricted to two dimensional two-phase flows.

\section*{Acknowledgement}

The first author would like to acknowledge the fellowship received from Indian Institute of Technology Bombay as a part of the IRCC Research Internship Award. The support received from the Institute of Technology, Nirma University
$-$ for sending the first author to carry out research at Indian
Institute of Technology Bombay $-$ is gratefully acknowledged.
  \bibliographystyle{elsarticle-num} 
  \bibliography{AiMuR_bib2}

\nocite{*}





\end{document}